\documentclass[preprint,showpacs,preprintnumbers,amsmath,amssymb]{revtex4}
\usepackage{amsfonts}
\usepackage{amssymb}
\usepackage{latexsym}
\usepackage{graphicx}

\newcommand{\al}{\alpha}
\newcommand{\be}{\beta}
\newcommand{\pa}{\partial}

\newcommand{\si}{\sigma}

\newcommand{\de}{\delta}

\newcommand{\De}{\Delta}
\newcommand{\na}{\nabla}

\newcommand{\tha}{\theta}

\newcommand{\rar}{\rightarrow}

\begin{document}

\preprint{M\'exico ICN-UNAM 07/02, \ November, 2002}
\title{$H_2^{+}$ molecular ion in a strong magnetic field:
ground state}
\author{J.C. \surname{L\'opez Vieyra}}
\email{vieyra@nuclecu.unam.mx}
\affiliation{}
\author{A.V.~Turbiner}
\altaffiliation[]{On leave of absence from the Institute for Theoretical
 and Experimental Physics, Moscow 117259, Russia}
\email{turbiner@nuclecu.unam.mx}
\affiliation{Instituto de Ciencias Nucleares, UNAM,
Apartado Postal 70-543, 04510 M\'exico}

\date{November 15, 2002}

\begin{abstract}
  A detailed quantitative analysis of the system $(ppe)$ placed in
  magnetic field ranging from $0 - 4.414 \times 10^{13}\,G$  is presented.
  The present study is focused on the question of the existence of the molecular
  ion $H_2^{+}$ in a magnetic field. As a tool, a variational method with an
  optimization of the form of the vector potential (optimal gauge fixing)
  is used.  It is shown that in the domain of applicability of the non-relativistic
  approximation the system $(ppe)$ in the Born-Oppenheimer approximation has a
  well-pronounced minimum in the total energy at a finite interproton distance
  for $B \lesssim 10^{11}\,G$, thus manifesting the existence
  of  $H_2^{+}$. For $B \gtrsim 10^{11}\,G$ and large inclinations (of the
  molecular axis with respect to the magnetic line) the minimum disappears and hence
  the molecular ion $H_2^{+}$ does not exist. It is shown that the most stable
  configuration of $H_2^{+}$ always corresponds to protons situated along the
  magnetic line. With magnetic field growth the ion $H_2^{+}$ becomes more and
  more tightly bound and compact, and the electronic distribution evolves
  from a two-peak to a one-peak pattern. The domain of inclinations where the
  $H_2^{+}$ ion exists reduces with magnetic field increase and finally becomes
  $0^o - 25^o$ at $B = 4.414 \times 10^{13}\,G$. Phase transition type behavior
  of variational parameters for some interproton distances related to the beginning
  of the chemical reaction $H_2^{+} \leftrightarrow H + p$ is found.
\end{abstract}


\maketitle

\section{Introduction}
\label{sec:intro}

Many years have passed since the moment when theoretical
qualitative arguments were given that show that in the presence of
a strong magnetic field the physics of atoms and molecules
exhibits a wealth of new, unexpected phenomena even for the
simplest systems \cite{Kadomtsev:1971,Ruderman:1971}. In
particular, a chance that unusual chemical compounds may be formed
which do not exist without magnetic field was mentioned. In
practice, the atmosphere of neutron stars, which is characterized
by the presence of enormous magnetic fields $10^{12} -
10^{13}\,G$, as well as other astronomical objects carrying large
magnetic fields ($>10^{8}\,G$) provide a valuable paradigm where
this physics could be realized. Recently, the experimental data
collected by the {\it Chandra} X-ray observatory revealed certain
irregularities in the spectrum of an isolated neutron star
1E1207.4-5209. These irregularities can be interpreted as
absorption features at $\sim 0.7\,$ KeV and $\sim 1.4\,$ KeV of
possible atomic or molecular nature \cite{Sandal:2002}.

One of the first general features observed in standard atomic and
molecular systems placed in a strong magnetic field is an increase
of both total and binding energies, accompanied by a drastic
shrinking of the electron localization length in both the
longitudinal and transverse directions. Naturally, this leads to a
decrease of the equilibrium distance with magnetic field growth.
This behavior can be considered to be a consequence of the fact
that for large magnetic fields the electron cloud takes a
needle-like form extended along the magnetic field direction and
the system becomes effectively quasi-one-dimensional
\cite{Ruderman:1971}. It is obvious that the phenomenon of
quasi-one-dimensionality enhances the stability of standard atomic
and molecular systems from the electrostatic point of view. In
particular, molecules become elongated along the magnetic line
forming a type of linear molecular polymer (for details see the
review articles \cite{Garstang:1977, Lai:2001}). It also hints at
the occurrence of exotic atomic and molecular systems which do not
exist in the absence of a magnetic field. Motivated by these
simple observations it was shown in Refs.\cite{Turbiner:1999,
Lopez-Tur:2000} that three and even four protons can be bound by
one electron. This shows that exotic one-electron molecular
systems $H_3^{2+}$ and $H_4^{3+}$ can exist in sufficiently strong
magnetic fields in the form of linear polymers. However, the
situation becomes much less clear (and also much less studied)
when the nuclei are not aligned with the magnetic field direction,
and thus in general do not form  a linear system. Obviously, such
a study would be important for understanding the kinetics of a gas
of molecules in the presence of a strong magnetic field. As a
first step towards such a study, even the simplest molecules in
different spatial configurations deserve attention. Recently, a
certain spatial configuration of $H_3^{2+}$ was studied in detail
\cite{Lopez-Tur:2002}. It was shown that in the range of magnetic
fields $10^8 < B < 10^{11}\,G$ the system $(pppe)$, with the
protons forming an equilateral triangle perpendicular to the
magnetic lines, has a well-pronounced minimum in the total energy
for a certain size of triangle. The goal of the present work is to
attempt for the first time to carry out an extensive quantitative
investigation of the ground state of $H_2^+$ in the framework of a
single approach in its entire complexity: a wide range of magnetic
field strengths ($0 - 4.414 \times 10^{13}\,G$), arbitrary (but
fixed) orientation of the molecular axis with respect to the
magnetic line and arbitrary internuclear distances. We are going
to carry out this study in the Born-Oppenheimer approximation at
zero order -- assuming protons to be infinitely heavy charged
centers. In principle, when the molecular axis is perpendicular to
the magnetic line the system $(ppe)$ acquires extra stability from
the electrostatic point of view. Electrostatic repulsion of the
classical protons is compensated for by the Lorentz force acting
on them.

It is well known that the molecular ion $H_2^+$ is the most stable
one-electron molecular system in the absence of a magnetic field.
It remains so in the presence of a constant magnetic field unless
$B \gtrsim 10^{13}\,G$, where the exotic ion $H_3^{2+}$ appears to
be the most bound (see \cite{Lopez-Tur:2000}). The ion $H_2^+$ has
been widely studied, both with and without the presence of a
magnetic field, due to its importance in astrophysics, atomic and
molecular physics, solid state and plasma physics (see
\cite{Garstang:1977}-\cite{Potekhin} and references therein). The
majority of the previous studies were focused on the case of the
parallel configuration, where the angle between the molecular axis
and the magnetic field direction is zero, $\tha=0^o$. The only
exception is Ref.\cite{Schmelcher}, where a detailed quantitative
analysis was performed for any $\tha$ but for a single magnetic
field $B=1\,a.u.\,$. Previous studies were based on various
numerical techniques, but the overwhelming majority used different
versions of the variational method, including the Thomas-Fermi
approach. As a rule, in these studies the nuclear motion was
separated from the electronic motion using the Born-Oppenheimer
approximation at zero order (see above). It was observed at the
quantitative level that magnetic field growth is always
accompanied by an increase in the total and binding energies, as
well as a shrinking of the equilibrium distance. As a consequence
it led to a striking conclusion about the drastic increase in the
probability of nuclear fusion for $H_2^+$ in the presence of a
strong magnetic field \cite{Kher}.

In the present study we will also use a variational method. Our
consideration will be limited by a study of the $1_g$-state, which
realizes the ground state of the system if the bound state exists
\footnote{After a straightforward separation of the spin part of
wavefunction, the original Schroedinger equation becomes a scalar
Schroedinger equation. Then it can be stated that a nodeless
eigenfunction corresponds to the ground state (Perron theorem)}.
We will construct state-of-the-art, non-straightforward,
`adequate' trial functions consistent with a variationally
optimized choice of vector potential. We should stress that a
proper choice of the form of the vector potential is one of the
crucial points which guarantee the adequacy and reliability of the
consideration. In particular, a proper position of the gauge
origin, where the vector potential vanishes, is drastically
important, especially for large interproton distances. For the
parallel configuration, $\tha=0^o$ the present work can be
considered as an extension (and also an development) of our
previous work \cite{Lopez:1997}. It is necessary to emphasize that
we encounter several new physical phenomena which occur when the
molecular axis deviates from the magnetic field direction. If the
magnetic field is sufficiently strong, $B\gtrsim 10^{11}\,G$ and
inclination $\tha$ is larger than a certain critical angle the ion
$H_2^+$ does not exist in the contrary to a prediction in
Refs.\cite{Larsen}, \cite{Kher}, \cite{Turbiner:2001}. This
prediction was based on an improper gauge dependence of the trial
functions, which caused a significant loss of accuracy and finally
led to a qualitatively incorrect result. We find that in the weak
field regime the $(ppe)$ system in the equilibrium position at any
inclination, the electronic distribution peaks at the positions of
the protons. While at large magnetic fields the electronic
distribution is characterized by single peak at the midpoint
between two protons. This change from a two-peak to a one-peak
configuration appears around $B \sim 10^{10}-10^{11}\,G$ with a
slight dependence on the inclination angle $\tha$. From a physical
point of view the former means that the electron prefers to stay
in the vicinity of a proton. This can be interpreted as a
dominance of the $H$-atom plus proton configuration. The latter
situation implies that the electron is `shared' by both protons
and hence such a separation to $H$-atom plus proton is irrelevant.
Therefore, we can call the two-peak situation `ionic' coupling,
while the one-peak case can be designated as `covalent' coupling,
although this definition differs from that widely accepted in
textbooks (see, for example \cite{LL}). Thus, we can conclude that
a new phenomenon appears - as the magnetic field grows the type of
coupling changes from `ionic' to `covalent'. At large internuclear
distances the electron is always attached to one of the charged
centers, so the coupling is `ionic'.

One particular goal of our study is to investigate a process of
dissociation of the $(ppe)$ system: $H_2^+ \rar H + p$ which
appears with increase of interproton distance. It is clear from a
physical point of view that at large distances the electronic
distribution should be first of the two-peak type and then should
change at asymptotically large distances, to a single-peak one,
but with a peak at the position of one of the protons. Somehow
this process breaks permutation symmetry and we are not aware of
any attempt to describe it. In our analysis this phenomenon
appears as a consequence of a change of position of the gauge
origin with increase of interproton distance.

From the physical point of view it is quite interesting to note
how the $(ppe)$ system behaves at very large interproton
distances. This domain is modelled by an $H$-atom plus proton
interaction. The interaction corresponds to
(magnetic-field-inspired-quadrupole) + charge interaction and is
dominant comparing to the standard Van der Waals force. For small
inclinations the above interaction is attractive as in the Van der
Waals case, but becomes {\it repulsive} for large inclinations.
This implies that the potential curves approach to asymptotic
value of total energy at the large interproton distances from
above contrary to the Van der Waals case.

The Hamiltonian which describes two infinitely heavy protons and
one electron placed in a uniform constant magnetic field directed
along the $z-$axis, ${\bf B}=(0,0,B)$ is given by (see e.g.
\cite{LL})
\begin{equation}
\label{Ham}
 {\cal H} = {\hat p}^2 + \frac{2}{R} -\frac{2}{r_1} -\frac{2}{r_2}
  + ({\hat p} {\cal A}+{\cal A}{\hat p}) +  {\cal A}^2 \ ,
\end{equation}
(see Fig.1 for notations), where ${\hat p}=-i \na$ is the
momentum, ${\cal A}$ is a vector potential, which corresponds to
the magnetic field $\bf B$. Hence, the total energy $E_T $ of
$H_2^+$ is defined as the total electronic energy plus the Coulomb
energy of proton repulsion. The binding energy is defined as an
affinity to have the electron at infinity, $E_b=B-E_T$. The
dissociation energy is defined as affinity to have a proton at
infinity, $E_d=E_H-E_T$, where $E_H$ is the total energy of the
hydrogen atom in magnetic field $B$.

\begin{figure}[tb]
\begin{center}
     \includegraphics*[width=4.5in]{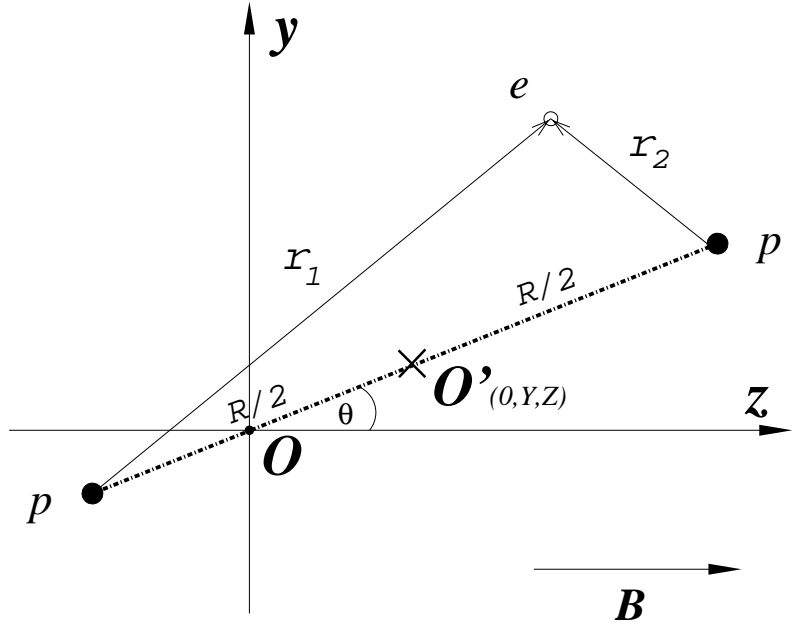}
     \caption{Geometrical setting  for the $H_2^{+}$ ion placed in a
      magnetic field directed along the $z$-axis. The protons are
      situated in the $y-z$ plane at a distance $R$ from each other
      and marked by bullets. $O$ is the origin of coordinates which is
      chosen to be on the bold-dashed line which connects protons;
      $O'(0,Y,Z)$ is the mid-point between protons. It is assumed
      that the gauge center coincides with $O$. $OO'$ measures the distance
      between the gauge center and the mid-point between the proton positions
      (see text and eq.(4)).}
    \label{fig:1}
\end{center}
\end{figure}

Atomic units are used throughout ($\hbar$=$m_e$=$e$=1) albeit
energies are expressed in Rydbergs (Ry). Sometimes, the magnetic
field $B$ is given in a.u. with $B_0= 2.35 \times 10^9\,G$
 \footnote{In absence of convention, some results presented in
literature are obtained for $B_0= 2.3505 \times 10^9\,G$. Thus,
making a comparison of the results obtained by different authors,
this fact should be taken into account}.

\section{Optimization of vector potential}

It is well known that the vector potential for a given magnetic
field, even in the Coulomb gauge $(\na \cdot {\cal A}) =0$, is
defined ambiguously, up to a gradient of an arbitrary function.
This gives rise a feature of gauge invariance: the Hermitian
Hamiltonian is gauge-covariant, while the eigenenergies and other
observables are gauge-independent. However, since we are going to
use an approximate method for solving the Schroedinger equation
with the Hamiltonian (\ref{Ham}), our approximation of
eigenenergies can well be gauge-dependent (only the exact ones are
gauge-independent). Hence one can choose the form of the vector
potential in a certain optimal way. In particular, if the
variational method is used, the vector potential can be considered
as a variational function and be chosen by a procedure of
minimization.

Let us consider a certain one-parameter family of vector
potentials corresponding to a constant magnetic field ${\bf
B}=(0,0,B)$
\begin{equation}
\label{Vec}
  {\cal A}= B((\xi-1)y,\ \xi x,\ 0)\ ,
\end{equation}
where $\xi$ is parameter, in the Coulomb gauge. The position of
the {\it gauge center} or {\it gauge origin}, where ${\cal
A}(x,y,z)=0$, is defined by $x=y=0$, with $z$ arbitrary. For
simplicity we fix $z=0$. If $\xi=1/2$ we get the well-known and
widely used gauge which is called symmetric or circular. If
$\xi=0$ or 1, we get the asymmetric or Landau gauge (see
\cite{LL}). By substituting (\ref{Vec}) into (\ref{Ham}) we arrive
at a Hamiltonian of the form
\begin{widetext}
\begin{eqnarray}
  \label{Ham.fin}
 {\cal H} = -{\nabla}^2 + \frac{2}{ R} -\frac{2}{r_1}
 -\frac{2}{r_2}
 -  2 i B[(\xi-1) y \pa_x  + \xi x \pa_y]
+  B^2 [ \xi^2 x^2+ (1-\xi)^2 y^2 ] \ ,
\end{eqnarray}
\end{widetext}
where $R$ is the interproton distance (see Fig. 1).

It is evident that for small interproton distances, $R$, the
electron prefers to be near the mid-point between the two protons
(coherent interaction with the protons). In the opposite limit,
$R$ large, the electron is situated near one of the protons (this
is an incoherent situation - the electron selects and then
interacts essentially with one proton). This fact, together with
naive symmetry arguments, leads us to a natural assumption that
the gauge center is situated on a line connecting the protons.
Therefore, the coordinates of mid-point between protons are
\begin{equation}
\label{d}
 Y=\frac{Rd}{2} \sin \tha \ ,\ Z=\frac{Rd}{2} \cos \tha \ ,
\end{equation}
(see Fig.1), where $d$ is a parameter. Thus, the position of the
gauge center is effectively measured by the parameter $d$ -- a
relative distance between the middle of the line connecting the
protons and the gauge center. If the mid-point coincides with the
gauge center, then $d=0$. On other hand, if the position of a
proton coincides with the gauge center, then $d=1$ or $d=-1$.
Hence the parameter $d$ makes sense as a parameter characterizing
a gauge.

The idea of choosing an optimal (convenient) gauge has been widely
exploited in quantum field theory calculations. It has also been
discussed in quantum mechanics and, in particular, in connection
to the present problem. Perhaps, the first constructive (and
remarkable) attempt to realize the idea of an optimal gauge was
made in the eighties by Larsen \cite{Larsen}. In his variational
study of the ground state of the $H_2^+$ molecular ion it was
explicitly shown that for a given fixed trial function the gauge
dependence of the energy can be quite significant. Furthermore,
even an oversimplified optimization procedure improves the
accuracy of the numerical results \footnote{For review of
different optimization procedures of vector potential see, for
instance, \cite{Kappes:1994} and references therein}.

Our present aim is to study the ground state of (\ref{Ham}) or,
more concretely, (\ref{Ham.fin}). We propose a different way of
optimization of vector potential than those discussed by previous
authors. It can be easily demonstrated that for a one-electron
system there always exists a certain gauge for which the ground
state eigenfunction is a real function. Let us fix a vector
potential in (\ref{Ham}). Assume that we have solved the spectral
problem exactly and have found the ground state eigenfunction. In
general, it is a certain {\it complex} function with a
non-trivial, coordinate-dependent phase. Treating this phase as a
gauge phase and then gauging it away, finally results in a new
vector potential. This vector potential has the property we want
-- the ground state eigenfunction of the Hamiltonian (\ref{Ham})
is real. It is obvious that similar considerations are valid for
any excited state. In general, for a given eigenstate there exists
a certain gauge in which the eigenfunction is real. For different
eigenstates these gauges can be different. It is obvious that
similar situation occurs for any one-electron system in a magnetic
field.

Dealing with real trial functions has an immediate advantage: the
expectation value of the terms proportional to ${\cal A}$ in
(\ref{Ham}) (or $\sim B$ in (\ref{Ham.fin})) vanishes when it is
taken over any real, normalizable function. Thus, without loss of
generality, the term $\sim B$ in (\ref{Ham.fin}) can be omitted.
Thus, we can use real trial functions with explicit dependence on
the gauge parameters $\xi$ and $d$. {\em These parameters are
fixed by performing a variational optimization of the energy}.
Therefore, as a result of the minimization we find both a
variational energy and a gauge for which the ground state
eigenfunction is real, as well as the corresponding Hamiltonian.
One can easily show that for a system possessing axial
(rotational) symmetry
 \footnote{ This is the case whenever the magnetic field is directed
 along the molecular axis (parallel configuration)}
the optimal gauge is the symmetric gauge $\xi=1/2$ with arbitrary
$d$. This is precisely the gauge which has been overwhelmingly
used (without any explanations) in the majority of the previous
research on $H_2^+$ in the parallel configuration
\cite{Kadomtsev:1971} - \cite{Potekhin}. However, this is not the
case if $\tha \neq 0^{\circ}$.  For the symmetric gauge the exact
eigenfunction now becomes complex, therefore complex trial
functions must be used. But following the recipe proposed above we
can avoid complex trial functions by adjusting the gauge in such a
way the eigenfunction remains real. This justifies the use of real
trial functions. Our results (see Section IV) lead to the
conclusion that for the ground state the optimal gauge parameter
varies in the interval $\xi \in [0.5\,,\,1]$.

\section{Choosing trial functions}

The choice of trial functions contains two important ingredients:
(i) a search for the gauge leading to the real, exact ground state
eigenfunction and (ii) performance of a variational calculation
based on {\it real} trial functions. The main assumption is that a
gauge corresponding to a real, exact ground state eigenfunction is
of the type (\ref{Vec}) (or somehow is close to it)
 \footnote{It can be formulated as a problem
--
  for a fixed value of $B$ and a given inclination to find a
  gauge for which the ground state eigenfunction is real.}.
In other words, one can say that we are looking for a gauge of
type (\ref{Vec}) which admits the best possible approximation of
the ground state eigenfunction by real functions. Finally, in
regard to our problem, the following recipe of variational study
is used:
 \emph{As the first step, we construct an {\bf adequate} variational
  real trial function $\Psi_0$ \cite{Tur}, for which the
  potential $V_0=\frac{\De \Psi_0}{\Psi_0}$ reproduces the original
  potential near Coulomb singularities and at large distances,
  where $\xi$ and $d$ would appear as parameters. The trial
  function should support symmetries of an
  original problem. We then perform a minimization of the energy
  functional by treating the free parameters of the trial function and
  $\xi, d$ on the same footing.}
In particular, such an approach enables us to find the
\emph{optimal} form of the Hamiltonian as a function of $\xi, d$.

The Hamiltonian (\ref{Ham}) gives rise to different symmetry
properties depending on the orientation of the magnetic field with
respect to the internuclear axis. The most symmetric situation
corresponds to $\tha=0^{\circ}$, where invariance under
permutation of the (identical) charged centers $P:
(1\leftrightarrow 2)$ together with $P_z: (z \to -z)$ holds. Since
the angular momentum projection $\ell_z =m$ is conserved, $P_z$
accounts also for the degeneracy $m\to -m$. Thus, we classify the
states as $ 1\si_{g,u}, 2\si_{g,u}, \ldots 1\pi_{g,u}, 2\pi_{g,u}
\ldots 1\de_{g,u}, 2\de_{g,u} \ldots$, where the numbers
$1,2,\ldots$ refer to the electronic states in increasing order of
energy. The labels $\si,\pi,\de \ldots$ are used to denote
$|m|=0,1,2 \ldots$, respectively, the label $g$ ($u$) gerade
(ungerade) is assigned to the states of even (odd) parity $P$ of
the system. At $\tha=90^o$ the Hamiltonian still remains invariant
under the parity operations $P$ and $P_z$, while the angular
momentum projection is no longer conserved and $m$ is no longer a
quantum number. The classification in this case is $1_{g,u}^\pm,
2_{g,u}^\pm, \ldots$, where the sign $+(-)$ is used to denote even
(odd) $z$-parity. Eventually, for arbitrary orientation, only
parity under permutations $P$ is conserved. In general, we refer
to the lowest gerade and ungerade states in our study as $1_g$ and
$1_u$.  This is the only unified notation which make sense for all
orientations $0^{\circ} \leq \tha \leq 90^{\circ}$.

The above recipe (for the symmetric gauge where $\xi=1/2, d=0$)
was successfully applied in a study of the $H_2^+$-ion in a
magnetic field for the parallel configuration $\tha=0^o$
\cite{Lopez:1997} and also for general one-electron linear systems
aligned along the magnetic field \cite{Lopez-Tur:2000}. In
particular, this led to the prediction of the existence of the
exotic ions $H_3^{2+}$ at $B \gtrsim 10^{10}\,G$ and in a linear
configuration $H_4^{3+}$ at $B \gtrsim 10^{13}\,G$
\cite{Turbiner:1999, Lopez-Tur:2000}. Recently, this recipe was
used for the first time to make a detailed study of the spatial
configuration $H_3^{2+}$ \cite{Lopez-Tur:2002}. It was
demonstrated that inconsistency between the form of vector
potential and a choice of trial functions can lead to non-trivial
artifacts like existence of spurious bound states (see
\cite{Lopez:2000}).

One of the simplest trial functions for $1_g$ state which meets
the requirements of our criterion of adequacy is
\begin{eqnarray}
\label{tr:1}
 \Psi_1= {e}^{-\al_1  (r_1+r_2)}
 {e}^{- B  [\be_{1x} \xi x^2 + \be_{1y}(1-\xi) y^2] }\,
\end{eqnarray}
(cf. \cite{Lopez:1997, Turbiner:2001}), where $\al_1$, $\be_{1x}$
and $\be_{1y}$ are variational parameters and $\xi$ is the
parameter of the gauge (2).  The first factor in the function
(\ref{tr:1}), being symmetric under interchange of the charge
centers $r_1\leftrightarrow r_2$, corresponds to the product of
two $1s$-Coulomb orbitals centered on each proton. It is nothing
but the celebrated Heitler-London approximation for the ground
state $1\si_g$. The second factor is the lowest Landau orbital
corresponding to the vector potential of the form Eq. (\ref{Vec}).
So, the function (\ref{tr:1}) can be considered as a modification
of the free field Heitler-London function. Following the
experience gained in studies of $H_2^+$ without a magnetic field
it is natural to assume that Eq. (\ref{tr:1}) is adequate to
describe interproton distances near equilibrium. This assumption
will be checked (and eventually confirmed) {\it a posteriori},
after making concrete calculations (see Section IV).

The function (\ref{tr:1}) is an exact eigenfunction in the
potential
\[
  V_1 =\frac{\nabla^2 \Psi_1}{\Psi_1} =
2 \al_1^2 - 2B[\be_{1x}\xi+\be_{1y}(1-\xi)] + 4 B^2 [ \be_{1x}^2
\xi^2 x^2 + \be_{1y}^2(1-\xi)^2 y^2 ] + 2 \al_1^2 ({\hat
n_1}\cdot{\hat n_2})
\]
\[
 + 4 {\al_1} B \left[
\frac{\be_{1x}\xi x^2 + \be_{1y}(1-\xi) y (y-y_1)}{r_1} +
\frac{\be_{1x}\xi x^2 + \be_{1y}(1-\xi) y (y-y_2)}{r_2} \right]
 -2 \al_1 \left[ \frac{1}{r_1} + \frac{1}{r_2}
\right]\ ,
\]
where $y_{1,2}$ are the $y$-coordinates of protons (see Fig.1).
The potential $V_1$ reproduces the functional behavior of the
original potential (\ref{Ham.fin}) near Coulombic singularities
and at large distances. These singularities are reproduced exactly
when $\be_{1x}=\be_{1y} =1/2$ and $\al_1=1$.

One can construct another trial function which meets the
requirements of our criterion of adequacy as well,
\begin{equation}
\label{tr:2} \Psi_2= \bigg({e}^{-\al_2 r_1} + {e}^{-\al_2
r_2}\bigg) {e}^{ -  B  [\be_{2x}\xi x^2 +\be_{2y}(1-\xi) y^2] }\ ,
\end{equation}
(cf. \cite{Lopez:1997, Turbiner:2001}).  This is the celebrated
Hund-Mulliken function of the free field case multiplied by the
lowest Landau orbital, where $\al_2$, $\be_{2x}$ and $\be_{2y}$
are variational parameters. From a physical point of view this
function has to describe the interaction between a hydrogen atom
and a proton (charge center), and, in particular, models the
possible dissociation mode of $H_2^+$ into a hydrogen atom plus
proton. Thus, one can naturally expect that for sufficiently large
internuclear distances $R$ this function prevails, giving a
dominant contribution. Again this assumption will be checked {\it
a posteriori}, by concrete calculations (see Section IV).

There are two natural ways to incorporate the behavior of the
system in both regimes -- near equilibrium and at large distances
-- into a single trial function. It is to make a linear or a
nonlinear interpolation. The linear interpolation is given by a
linear superposition
\begin{equation}
\label{tr:3a} \Psi_{3a}= A_1 \Psi_{1} + A_2 \Psi_{2}\ ,
\end{equation}
where $A_1$ or $A_2$ are parameters and one of them is kept fixed
by a normalization condition. In turn, the simplest nonlinear
interpolation is of the form
\begin{equation}
\label{tr:3b} \Psi_{3b}= \bigg({e}^{-\al_3 r_1-\al_4 r_2} +
{e}^{-\al_3 r_2-\al_4 r_1}\bigg) {e}^{ - B  [\be_{3x}\xi x^2
+\be_{3y}(1-\xi) y^2] }\ ,
\end{equation}
(cf. \cite{Lopez:1997, Turbiner:2001}), where $\al_{3}$,
$\al_{4}$, $\be_{3x}$ and $\be_{3y}$ are variational parameters.
This is a Guillemin-Zener function for the free field case
multiplied by the lowest Landau orbital. If $\al_3=\al_4$, the
function (\ref{tr:3b}) coincides with (\ref{tr:1}). If $\al_4=0$,
the function (\ref{tr:3b}) coincides with (\ref{tr:2}).

The most general Ansatz is a linear superposition of the trial
functions (\ref{tr:3a}) and (\ref{tr:3b}),
\begin{equation}
\label{trial}
 \Psi = A_1 \Psi_{1} + A_2 \Psi_{2}+A_{3}\Psi_{3b}\ ,
\end{equation}
where we fix one of the $A$'s and let all the other parameters
vary. Finally, the total number of variational parameters in
(\ref{trial}), including $R, \xi, d$, is fifteen for the ground
state. For the parallel configuration, $\tha=0^{\circ}$, the
parameters $\xi=0.5, d=0$ are fixed in advance and also
$\beta_{1x}=\beta_{1y}, \beta_{2x}=\beta_{2y},
\beta_{3x}=\beta_{3y}$. Hence the number of free parameters is
reduced to ten for the ground state.  Finally, with the function
(\ref{trial}) we intend to describe the ground state for {\it all}
magnetic fields where non-relativistic consideration is valid, $B
\leq 4.414 \times 10^{13}\,G$, and for {\it all} orientations of
the molecular axis.

Calculations were performed  using the minimization package MINUIT
from CERN-LIB. Numerical integrations were carried out with a
relative accuracy of $\sim 10^{-7}$ by use of the adaptive NAG-LIB
(D01FCF) routine. All calculations were performed on a PC
Pentium-III $800$ MHz.

\section{Results}

We carry out a variational study of the system $(ppe)$ with
infinitely heavy protons in the range of magnetic fields $0 < B <
4.414 \times 10^{13}\,G$, inclinations $0^o - 90^o$, for a wide
range of interproton distances. For magnetic fields $B <
10^{11}\,G$ the system displays a well-pronounced minimum in the
total energy at {\it all inclinations}. However, for $B >
10^{11}\,G$ at large inclinations the minimum in the total energy
disappears, while for small inclinations a minimum continues to
exist. This picture describes the domain of existence of the
molecular ion $H_2^+$. In general, we confirm a qualitative
results by Khersonskij \cite{Kher} about the {\it non-existence}
of a minimum at finite distances on the total energy surfaces of
the system $(ppe)$ at sufficiently strong magnetic fields for some
far from parallel orientations. It is worth mentioning that the
variational study in \cite{Kher} was carried out with a trial
function somewhat similar to that of Eq.(\ref{tr:2}), which,
however, does not fully fulfill our criterion of adequacy. The
potential corresponding to this function reproduces correctly the
original potential near Coulomb singularities and $\sim \rho^2$
growth at large distances. However, it generates growing terms
$\sim \rho$ which implies a reduction of the rate of convergence
of a perturbation theory for which the variational energy
represents the first two terms (see the discussion in \cite{Tur}).
Also, this trial function is not satisfactory from the point of
view of gauge invariance. However, in spite of all the
above-mentioned deficiencies it led to qualitatively correct
picture.

In Figs. 2-5 the total energy $E_T$ of the $(ppe)$ system as a
function of interproton distance $R$ for several values of the
magnetic field strength and different values of inclination $\tha$
is shown. For magnetic fields $B \lesssim 10^{11}\,G$ and for all
inclinations $0^o - 90^o$, each plot displays a well-pronounced
minimum at $R=R_{eq}$, manifesting the existence of the molecular
system $H_2^+$. For $B=1\,a.u.$ and $R \lesssim 3.5\,a.u.$ (see
Fig. 2) our results are similar to the results of
\cite{Wille:1988, Schmelcher} -- for fixed $R$ the potential
energy $E_T$ grows with inclination. In general, at large $R >
R_{eq}$ and for $\tha > 0^o$ all the curves behave alike: they
reveal a maximum and then tend (from above) to the total energy of
the hydrogen atom. For $\tha = 0^o$ the potential curves approach
to the asymptotics from below, displaying in general a behavior
similar to the field-free case, to a Van der Waals-force-inspired
behavior. This behavior is related to the fact that at large $R$
the configuration $H$-atom + proton appears. The $H$-atom has
quadrupole moment, $Q \sim B^2$ (see \cite{Tur, Lozovik,
Turbiner:1983, Potekhin}). Hence at large distances the total
energy is defined by a quadrupole moment - charge interaction
\begin{equation}
\label{quadrupole}
 E_T\ =\ -\frac{e Q(B) P_2(\cos \tha)}{R^3} \ ,
\end{equation}
where $P_2$ is the second Legendre polynomial. At small
inclinations $P_2(\cos \tha)$ is positive, the total energy is
negative, thus corresponding to attraction between the quadrupole
and the charge. Therefore, the total energy curve approaches to
the asymptotics from below. For large inclinations $P_2(\cos
\tha)$ is negative and the total energy is positive. Thus, this
corresponds to repulsion between quadrupole and charge, and
implies an existence of maximum of the total energy for large
interproton distances $R > R_{eq}$. We observe the maximum in all
Figs.2-5. It is worth mentioning that in the calculations
\cite{Wille:1988, Schmelcher} for $B=1\,a.u.$ and $\tha = 90^o$
(and other inclinations) the maximum was not observed in
contradiction to our predictions (see Fig. 2 and also below Fig.
9). Looking at Fig. 2 it is interesting to compare a rate with
which potential curves are approaching to the asymptotic total
energy at large $R$. This asymptotic energy is equal to the total
energy of the hydrogen atom, $E_H=-0.6623 Ry$, while
$E_T^{\tha=0^o}(R=8\,a.u.)=-0.6647$ (from below),
$E_T^{\tha=45^o}(R=8\,a.u.)=-0.6576$ (from above),
$E_T^{\tha=90^o}(R=8\,a.u.)=-0.6620$ (from above). Thus, any
deviation does not exceed $1\%$. There exists a different manner
of viewing these results. It can be treated as a demonstration of
the quality of our trial function (\ref{trial}) but for the
calculation of the total energy of the atom (!).

\begin{figure}[tb]
\begin{center}
   \includegraphics*[width=4in,angle=-90]{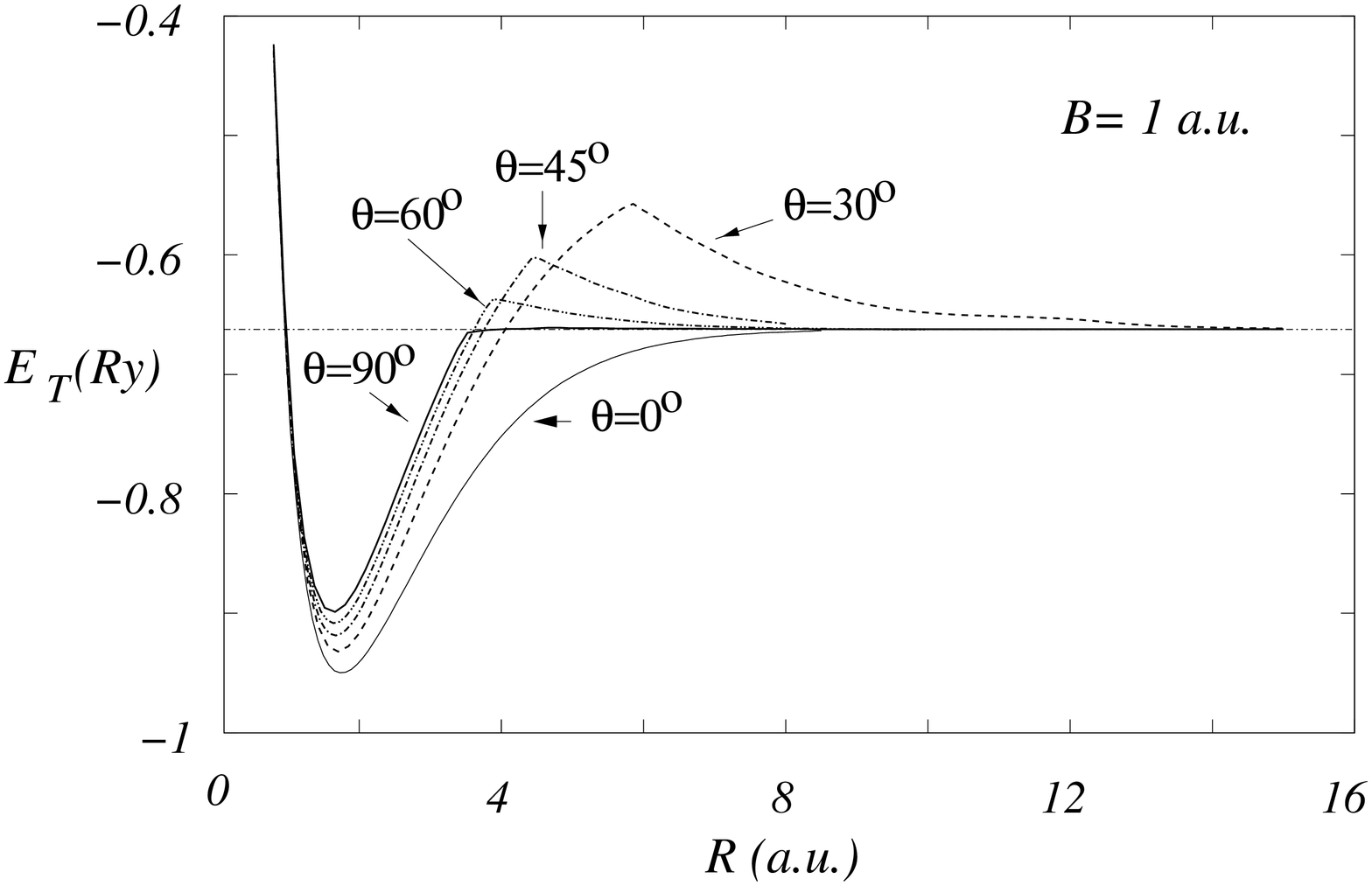}
   \caption{Total energy $E_T$ in Rydbergs of the $(ppe)$-system as function
   of interproton distance $R$ for different inclinations at
   $B = 2.3505 \times 10^9\,G \,(\,1 a.u.)$}
   \label{fig:2}
\end{center}
\end{figure}

\begin{figure}[tb]
\begin{center}
   \includegraphics*[width=4in,angle=-90]{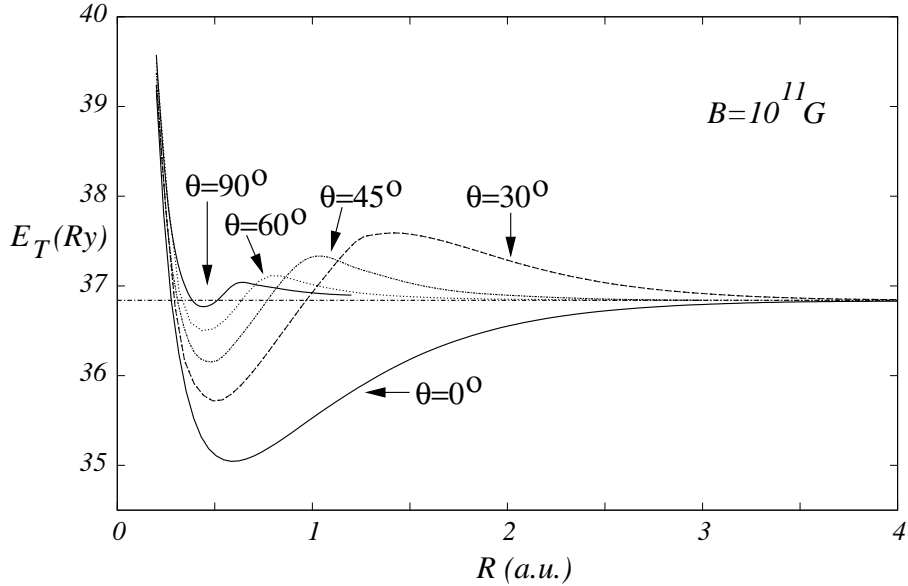}
   \caption{Total energy $E_T$ in Rydbergs of the $(ppe)$-system as function
   of interproton distance $R$ for different inclinations at $B = 10^{11}\,G$}
   \label{fig:3}
\end{center}
\end{figure}

\begin{figure}[tb]
\begin{center}
   \includegraphics*[width=4in,angle=-90]{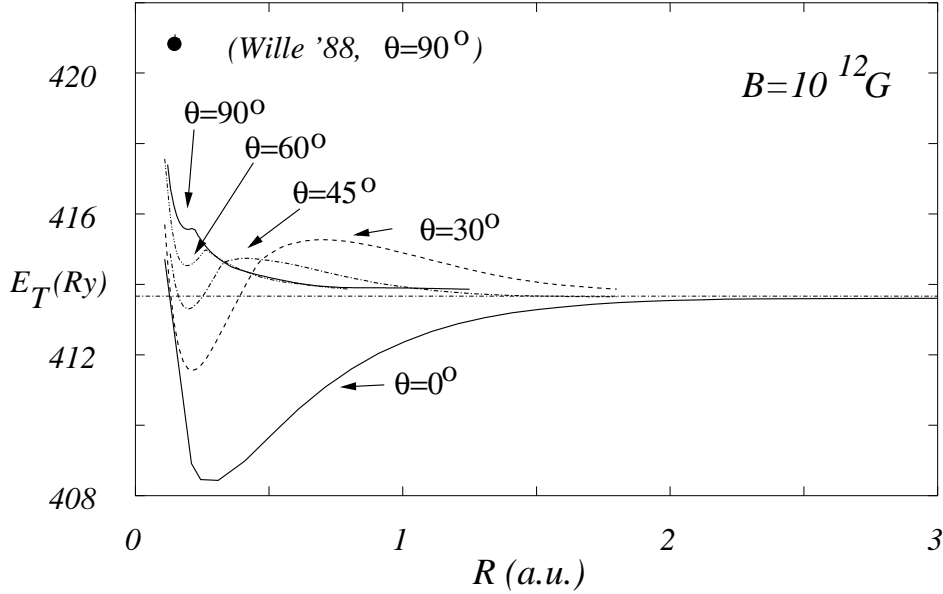}
   \caption{Total energy $E_T$ in Rydbergs of the $(ppe)$-system as function
   of interproton distance $R$ for different inclinations at $B = 10^{12}\,G$.
   The result by Wille \cite{Wille:1988} is shown by bullet
   (see text)}
   \label{fig:4}
\end{center}
\end{figure}

\begin{figure}[tb]
\begin{center}
   \includegraphics*[width=4in,angle=-90]{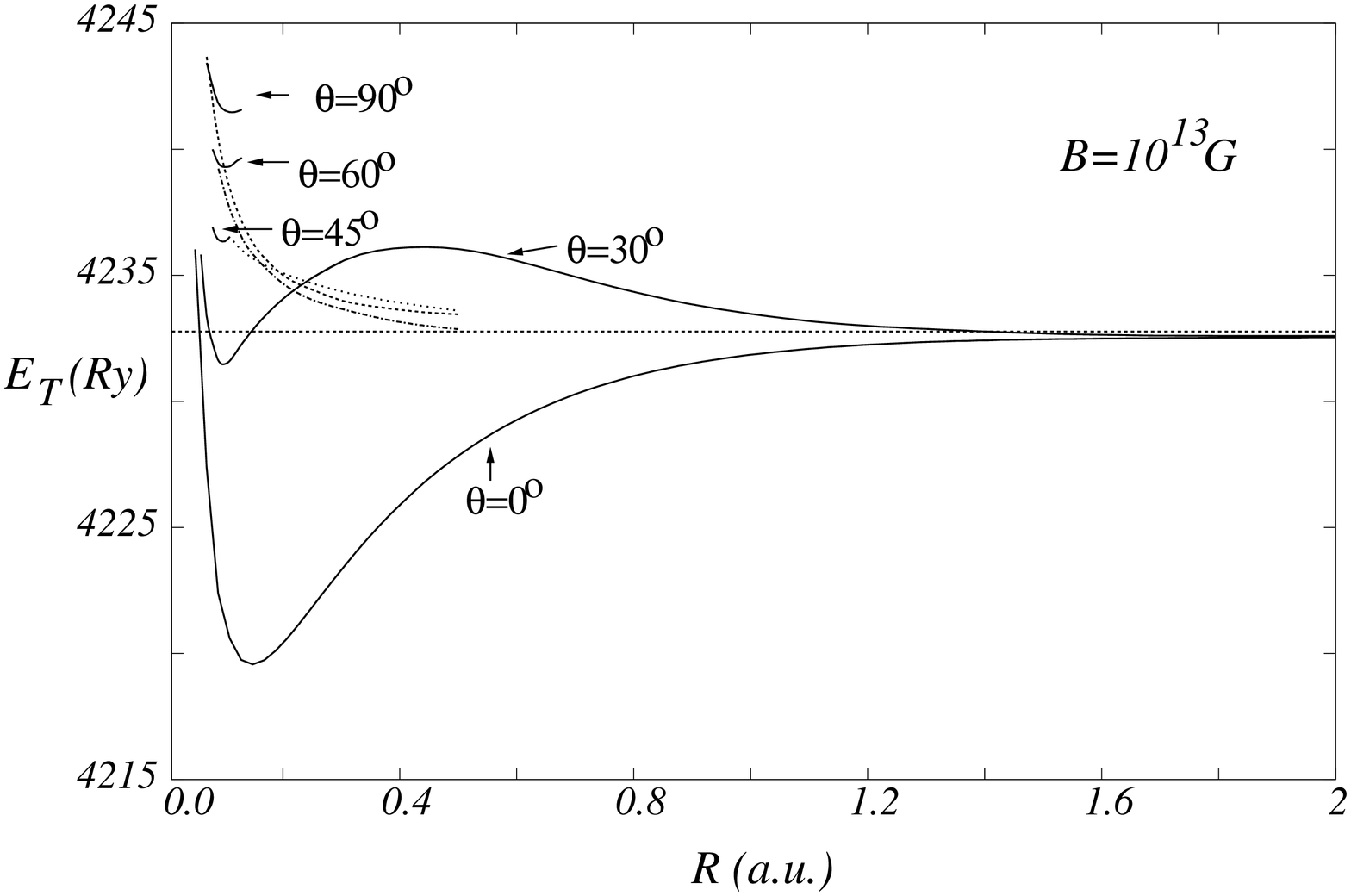}
   \caption{Total energy $E_T$ in Rydbergs of the $(ppe)$-system as a function
   of interproton distance $R$ for different inclinations at $B = 10^{13}\,G$.
   Plots for $\tha=45^o, 60^o, 90^o$ consist of two parts -- (i) solid
   line, when $d$ is kept fixed, $d=0$ (gauge center coincides with the
   mid-point between protons) and which displays a minimum, and
   (ii) the dotted line is the result of minimization, when the
   parameter $d$ is released.
   }
   \label{fig:5}
\end{center}
\end{figure}

However, the situation is drastically different for $B >
10^{11}\,G$, see Figs. 4-5. There exists a certain critical angle
$\tha_{cr}$, such that for $\tha < \tha_{cr}$ the situation
remains similar to that given above -- each potential curve is
characterized by a well-pronounced minimum at finite $R$. With
increase of the inclination, at $\tha \gtrsim \tha_{cr}$ the
minimum in the total energy first becomes very shallow with $E_T
> E_{H}$ and ceases to exist at all. We were unable to
localize with a confidence the domain in $R$ which correspond to a
shallow minimum which leads to the possible dissociation $H_2^+
\rar H + p\,$ that was predicted in \cite{Larsen} as well as being
discussed in our previous work \cite{Turbiner:2001}. We consider
that the prediction of dissociation for large inclinations emerged
as an artifact of an improper choice of the gauge fixing (see the
discussion above). A detailed study of the transition domain
(existence $\leftrightarrow$ non-existence) of $H_2^+$ is beyond
of scope of the present article. In any case, such a study
requires much more accurate quantitative techniques as well as a
sophisticated qualitative analysis. Schematically, the situation
is illustrated in Fig. 6.

\begin{figure}
\includegraphics*[width=3.5in,angle=-90]{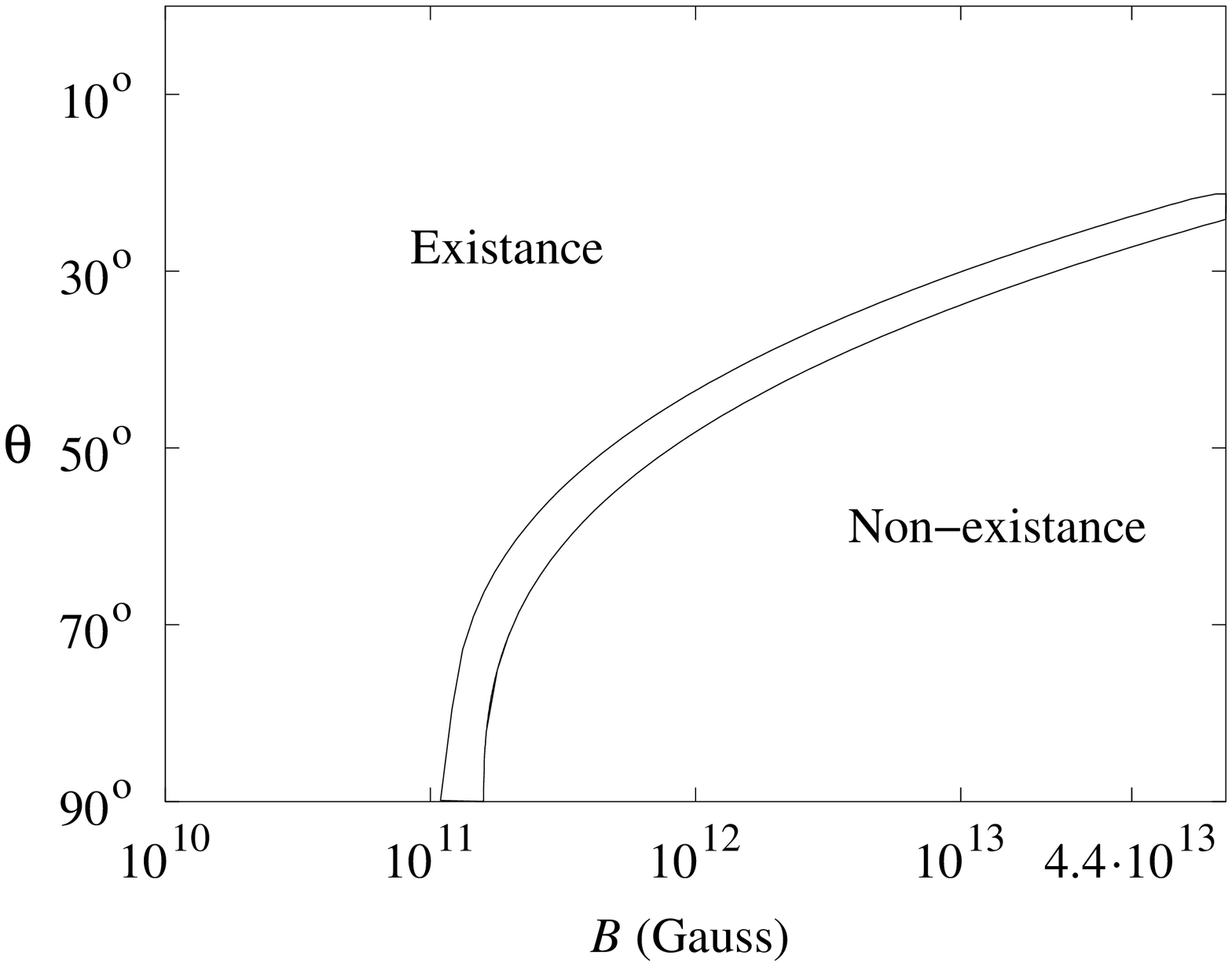}%
   \caption{\label{fig:6} $H_2^+$-ion: domains of existence
   $\leftrightarrow$ non-existence for the $1_g$ state.}
\end{figure}

\renewcommand{\thefigure}{7{\alph{figure}}}
\setcounter{figure}{0}

\begin{figure}[tb]
\begin{center}
   \includegraphics*[width=4in,angle=-90]{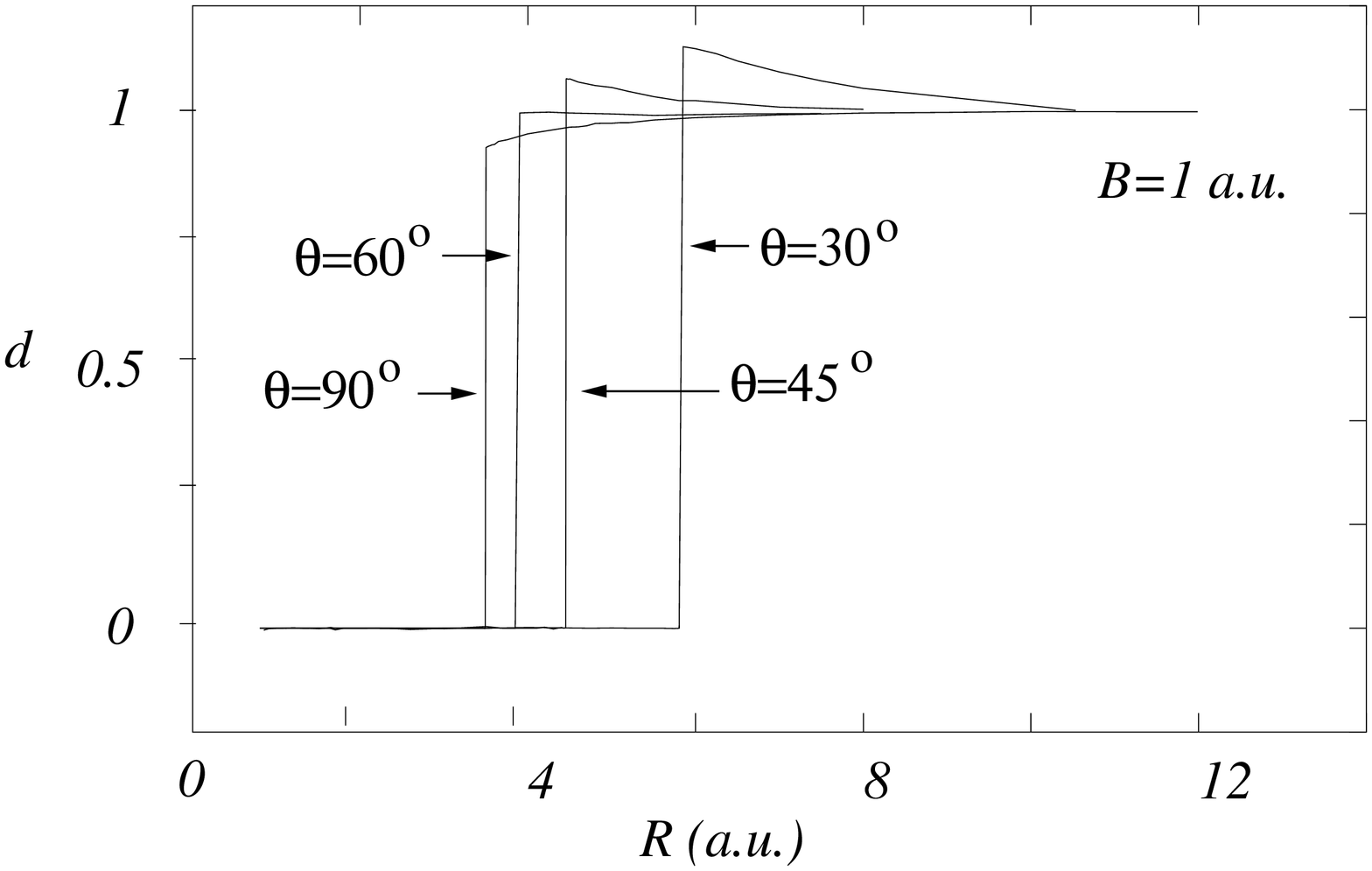}
   \caption{The dependence $d$ {\it vs} $R$ at $B=1\,a.u.$
   for different inclinations $\tha \neq 0^o$.}
   \label{fig:7a}
\end{center}
\end{figure}

\begin{figure}[tb]
\begin{center}
   \includegraphics*[width=4in,angle=-90]{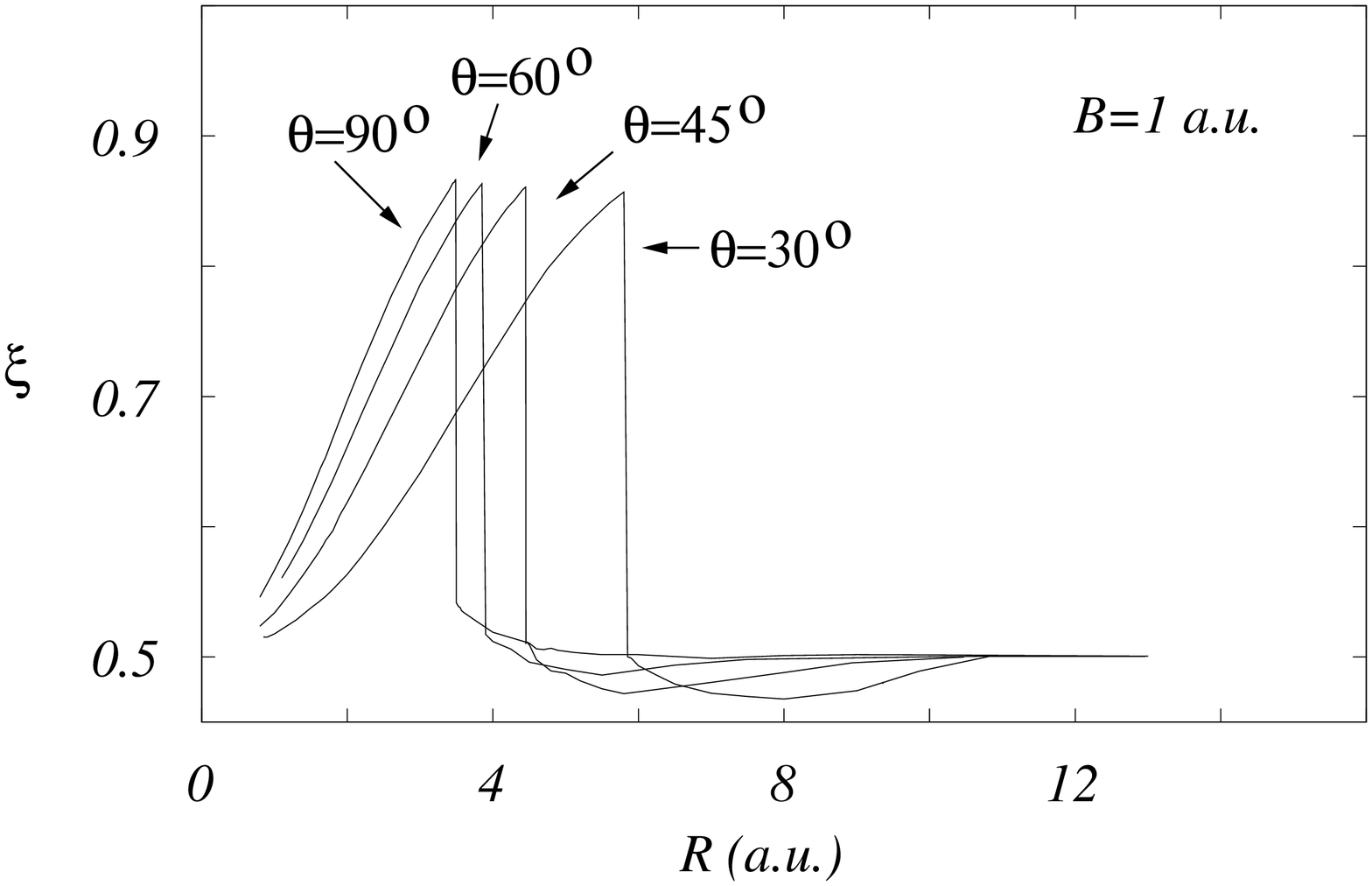}
   \caption{The dependence $\xi$ {\it vs} $R$ at $B=1\,a.u.$
   for different inclinations $\tha \neq 0^o$.}
   \label{fig:7b}
\end{center}
\end{figure}

\renewcommand{\thefigure}{8{\alph{figure}}}
\setcounter{figure}{0}

\begin{figure}[tb]
\begin{center}
   \includegraphics*[width=4in,angle=-90]{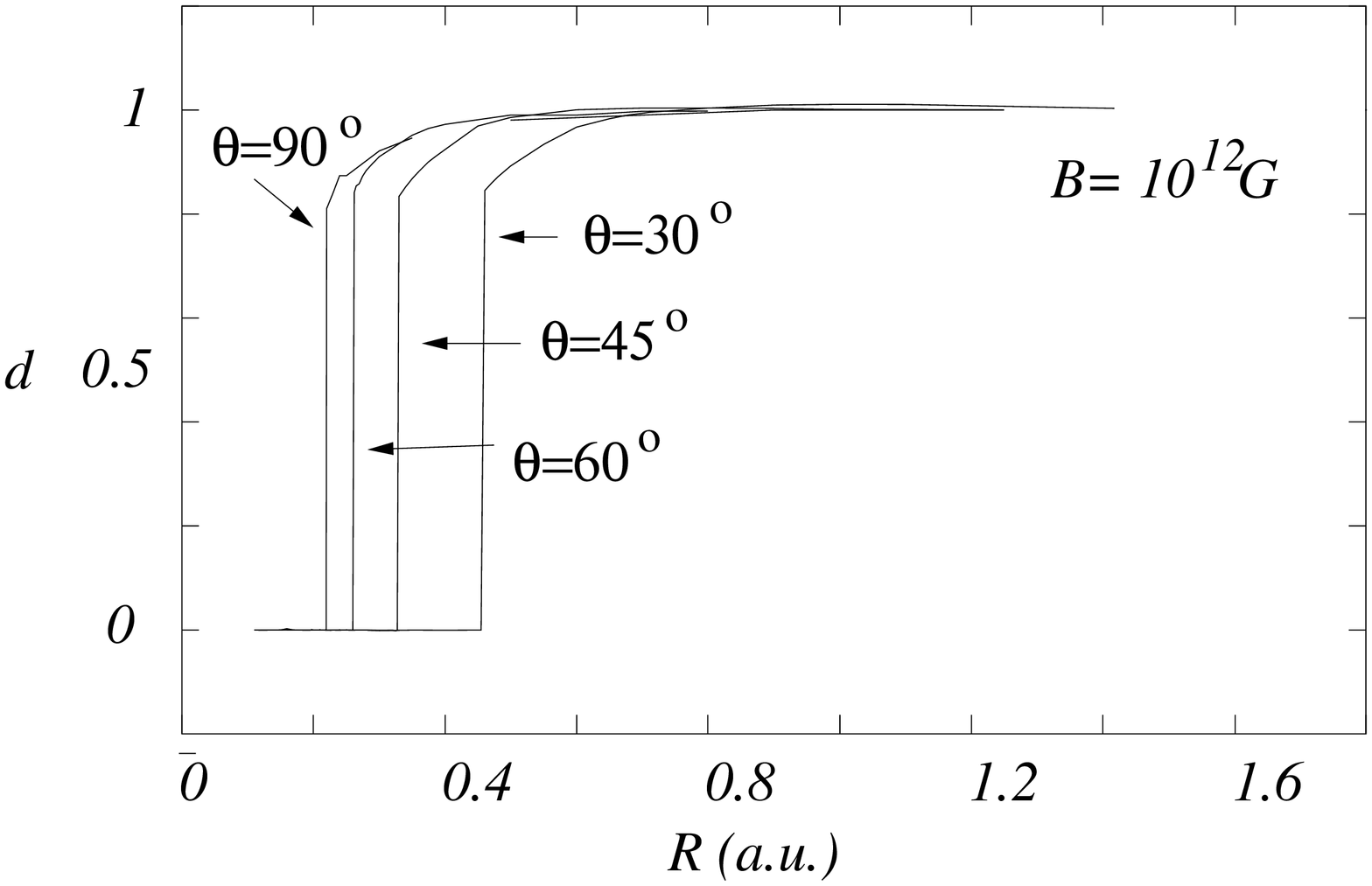}
   \caption{The dependence $d$ {\it vs} $R$ at $B=10^{12}\,G$
   for different inclinations $\tha \neq 0^o$.}
   \label{fig:8a}
\end{center}
\end{figure}

\begin{figure}[tb]
\begin{center}
   \includegraphics*[width=4in,angle=-90]{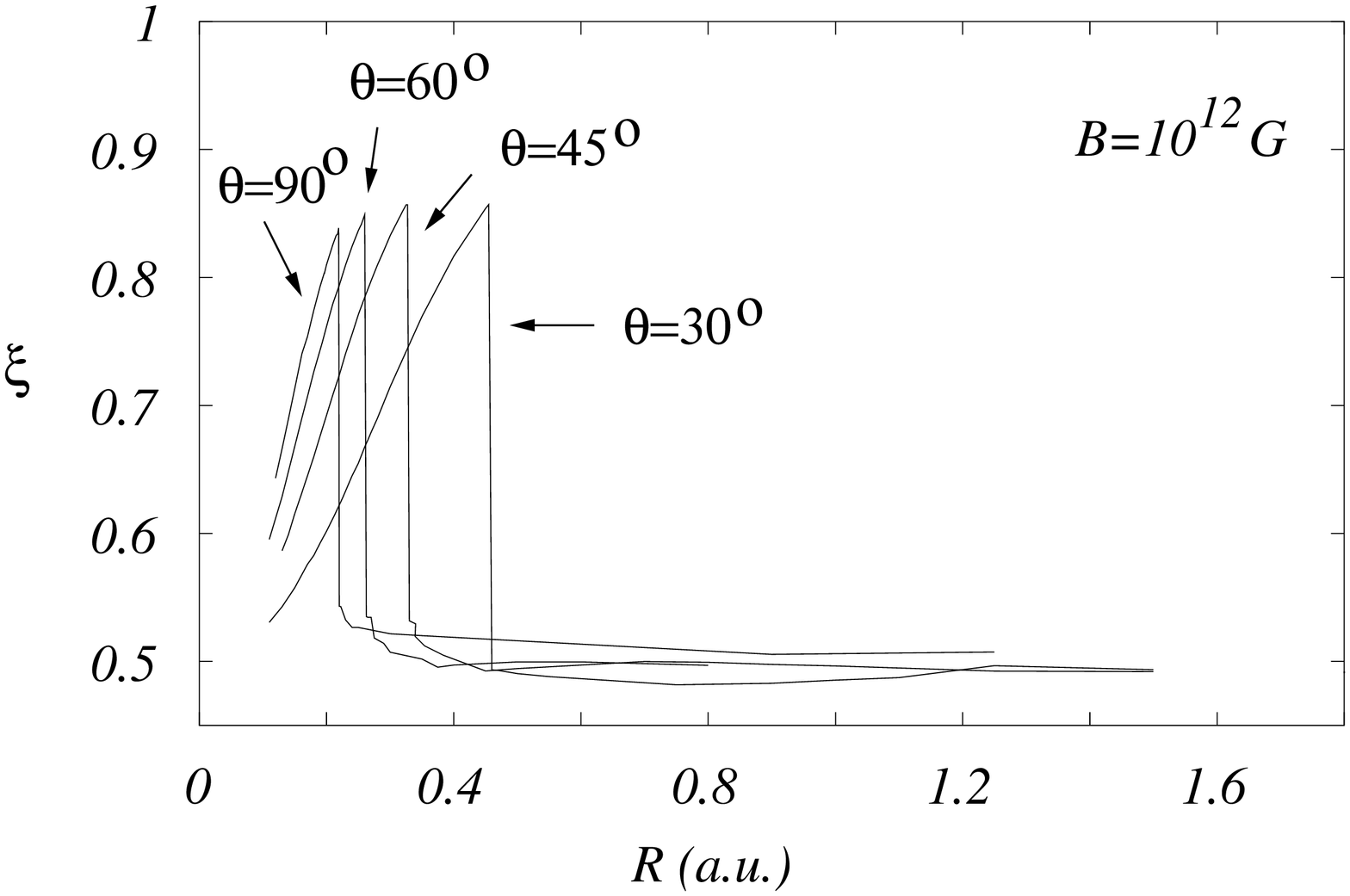}
   \caption{The dependence $\xi$ {\it vs} $R$ at $B=10^{12}\,G$
   for different inclinations $\tha \neq 0^o$.}
   \label{fig:8b}
\end{center}
\end{figure}

\renewcommand{\thefigure}{\arabic{figure}}
\setcounter{figure}{8}

It is quite interesting to explore the variation of the vector
potential (\ref{Vec}) for $\tha \neq 0^o$, in particular, the
position of the gauge center as a function of interproton distance
$R$ and magnetic field \footnote{At $\tha=0^o$ (parallel
configuration) the vector potential (\ref{Vec}) remains unchanged,
since $\xi=1/2$.}. In Figs. 7 a,b for $B=1a.u.$ and Figs. 8 a,b
for $10^{12}\,G$, correspondingly, both the $\xi-$ and $d-$
dependence are presented (see (\ref{Vec}) and discussion in
Sec.III). This dependence is very similar for all magnetic fields
studied. It is worthwhile to emphasize that for all the potential
curves given the minimum (in other words, the equilibrium
position) at $R=R_{eq}$ somehow corresponds to a gauge close to
the symmetric gauge: $\xi \gtrsim 1/2$ \footnote{The value of
$\xi$ grows with $B$ (see Figs.7a,b and below Tables II-III)} and
$Y=Z=0\,(d=0)$. A similar situation holds for small interproton
distances, $R<R_{eq}$. However, for large $R, R > R_{eq}$ the
parameter $\xi$ grows smoothly, reaching a maximum near the
maximum of the potential curve which we denote by $R=R_{cr}$. It
then falls sharply to the value $\xi \sim 1/2$. In turn, the
parameter $d$ remains equal 0 up to $R=R_{cr}$ (which means the
gauge center coincides to the mid-point between protons), then
sharply jumps to 1 (gauge center coincides with the position of a
proton), displaying a behavior similar to a phase transition. It
is indeed a type of {\it phase transition} behavior stemming from
symmetry breaking: from the domain $R<R_{cr}$, where the
permutation symmetry of the protons holds and where the protons
are indistinguishable, to the domain $R>R_{cr}$, where this
symmetry does not exist and electron is attached to one particular
proton. Such a type of `phase transitions' is typical in chemistry
and is called a `chemical reaction'. Hence the parameter $R_{cr}$
characterizes a distance at which the chemical reaction $H_2^+
\rar H + p$ starts.
Somewhat similar behavior of the gauge parameters has appeared in
the study of the exotic $H_3^{2+}$-ion \cite{Lopez-Tur:2002}.

\begin{figure}[tb]
\begin{center}
   \includegraphics*[width=4in,angle=-90]{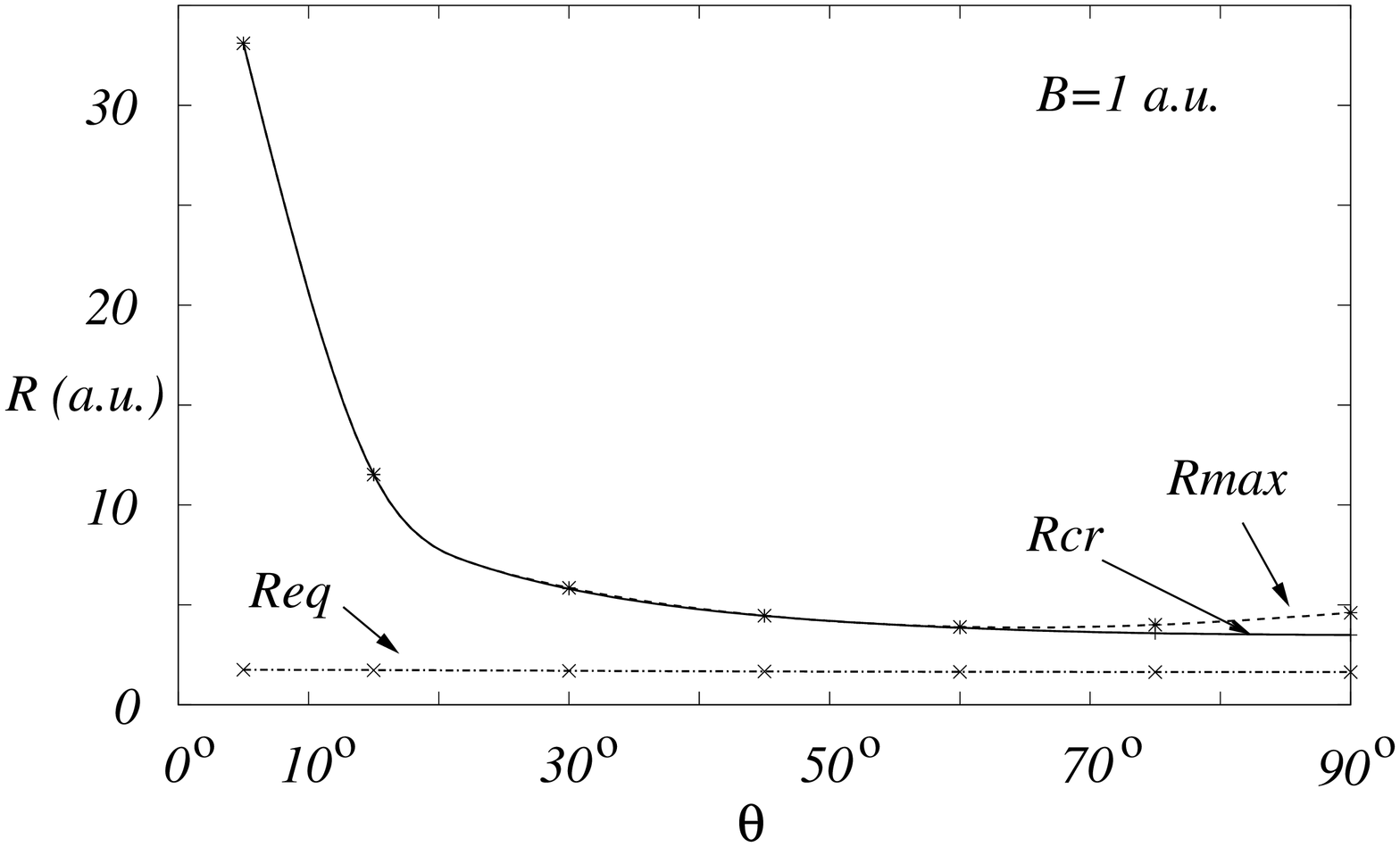}
   \caption{The dependence of $R_{crit}$ and position of the maximum
   $R_{max}$ compared to the equilibrium position $R_{eq}$
   at $B=1 a.u.$ for different inclinations $\tha$.}
   \label{fig:9}
\end{center}
\end{figure}

\begin{figure}[tb]
\begin{center}
   \includegraphics*[width=4in,angle=-90]{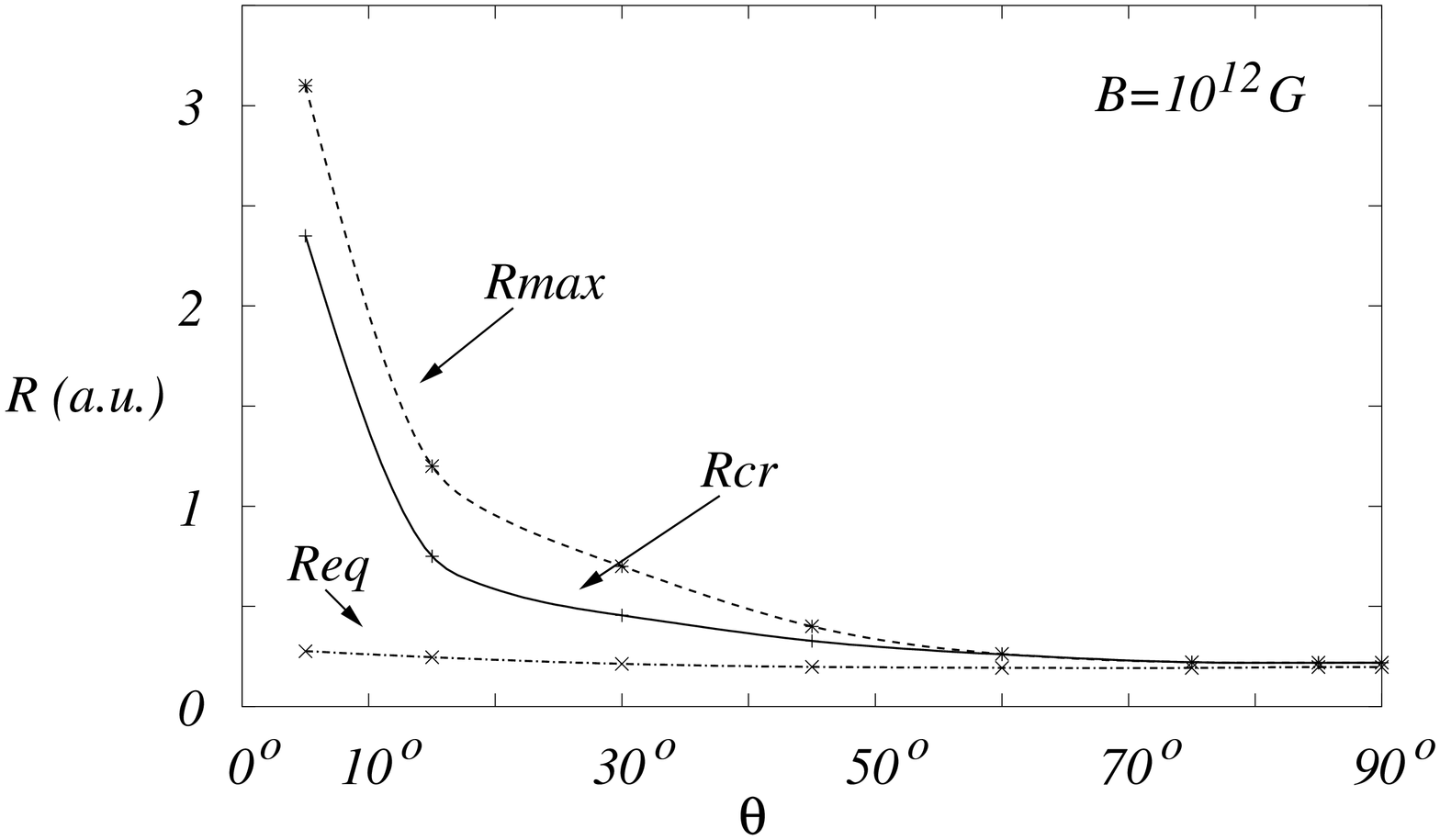}
   \caption{The dependence of $R_{crit}$ and position of the maximum
   $R_{max}$ compared to the equilibrium position $R_{eq}$
   at $B=10^{12}\,G$ for different inclinations $\tha$.}
   \label{fig:10}
\end{center}
\end{figure}

In Figs. 9-10 the behaviors of $R_{eq}, R_{max}, R_{cr}$ {\it vs}
inclination at $B=1 a.u.$ and $10^{12}\,G$ are displayed. The
behavior of $R_{eq}$ {\it vs} $\tha$ demonstrates almost no dependence
on $\tha$ in contrast to both $R_{max}$ and $R_{cr}$ which
drastically decrease with the growth of $\tha$. We do not have a
reliable physical explanation of this behavior.

The total energy dependence of $H_2^+$ (at $R=R_{eq}$) as a
function of the inclination angle $\tha$ for different magnetic
fields is shown in Fig. 11.  The dotted line corresponds to the
$H$-atom total energy in the corresponding magnetic field.  For
weak magnetic fields the hydrogen atom total energy is always
higher than that of the $H_2^+$-ion. However, for $B \gtrsim 2
\times 10^{11}\,G$ the situation changes -- a minimum of the
$H_2^+$ total energy for angles $\tha > \tha_{cr}$ does not exist
any more. Surprisingly, $\tha_{cr}$ corresponds approximately to
the moment when the total energy of the $H$-atom becomes equal to
the total energy of the $H_2^+$-ion. If the form of vector
potential (\ref{Vec}) is kept fixed with $\xi=1/2$ and
$Y=Z=0\,(d=0)$, then a spurious minimum appears; its position is
displayed by the dotted curve. However, if the gauge center
parameters are released this minimum disappears (see the
discussion above). It was an underlying reason for the erroneous
statement about the existence of the unstable $H_2^+$ ion in this
domain with a possibility to dissociate $H_2^+ \rar H + p$ (see
\cite{Turbiner:2001}). For all magnetic fields studied the total
energy is minimal at $\tha=0^o$ (parallel configuration) and then
increases monotonically with inclination in complete agreement
with statements of other authors \cite{Wille:1988, Kher, Larsen,
Schmelcher}.

\begin{figure*}
  \begin{center}
    \[
    \begin{array}{cc}
    {\includegraphics[width=2.4in,angle=-90]{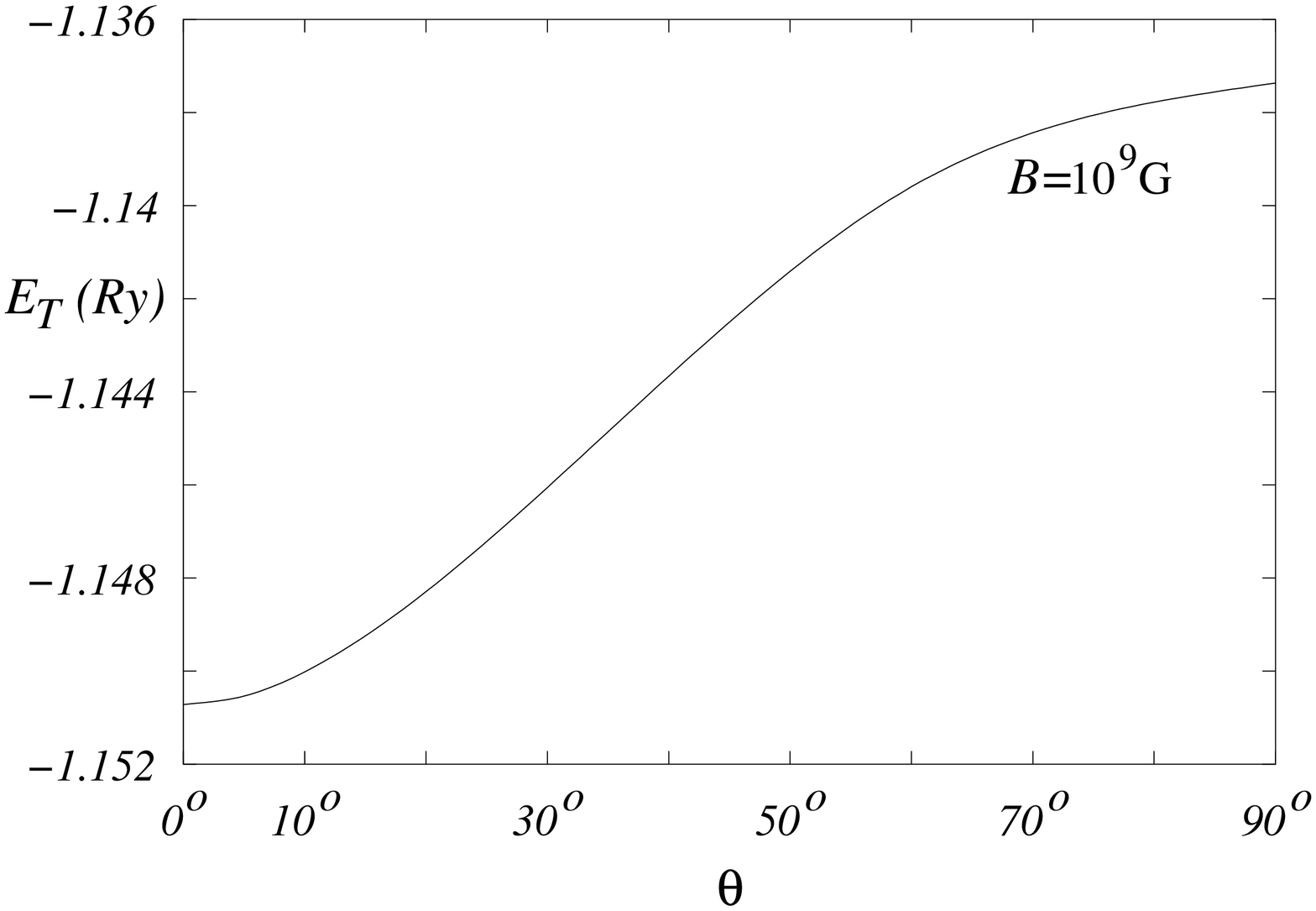}} &
    {\includegraphics[width=2.4in,angle=-90]{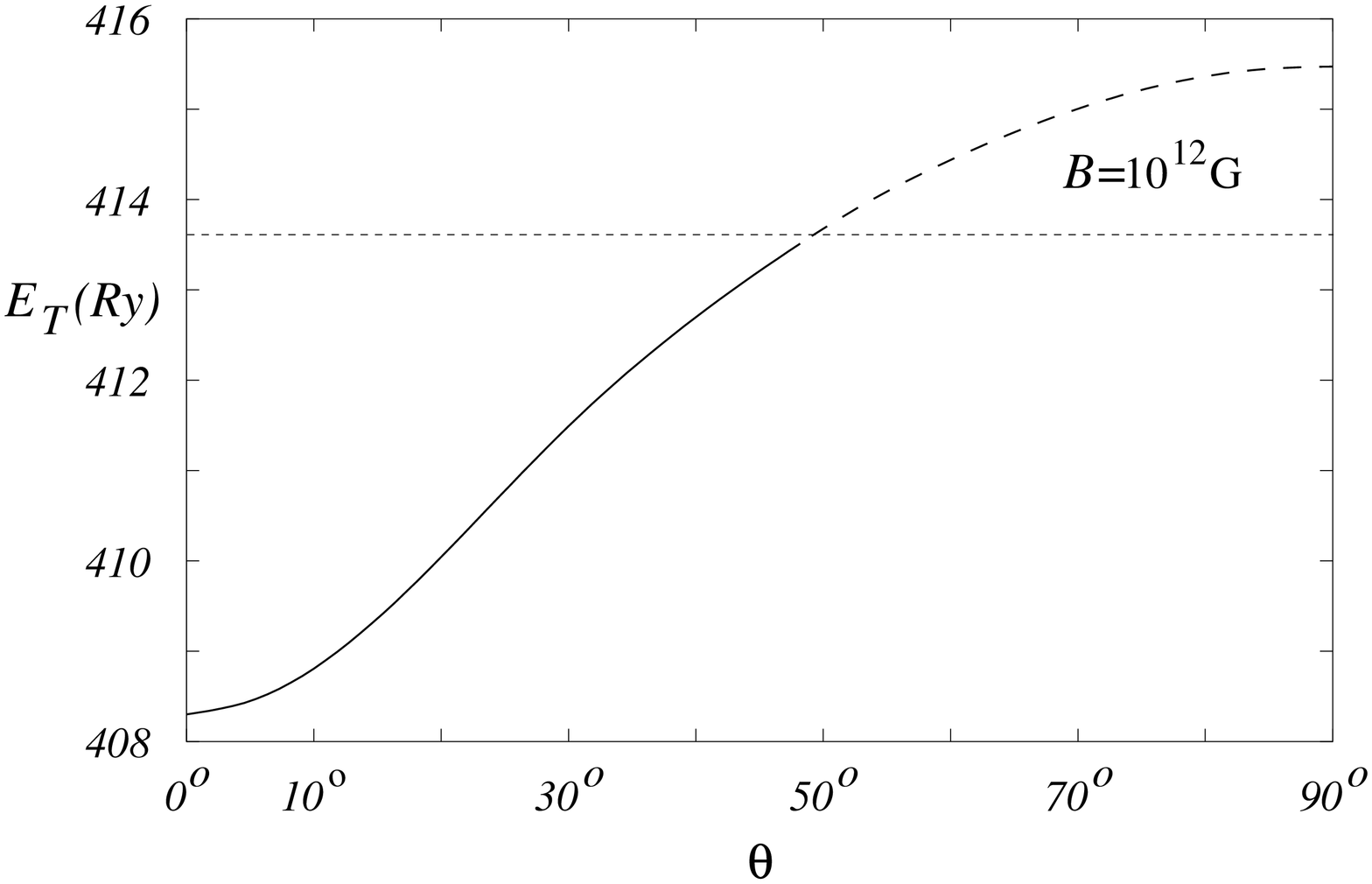}}     \\
    {\includegraphics[width=2.4in,angle=-90]{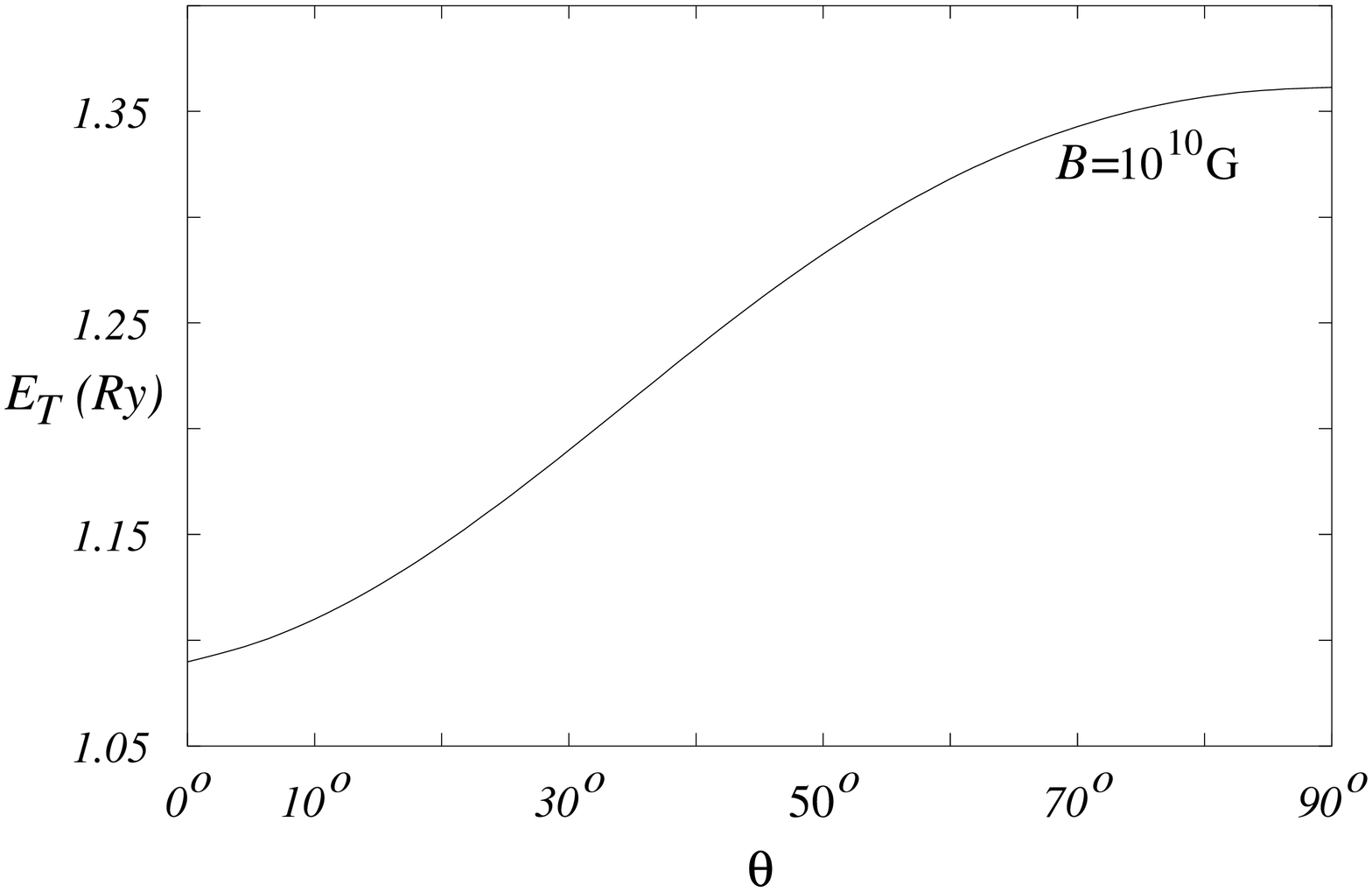}}&
    {\includegraphics[width=2.4in,angle=-90]{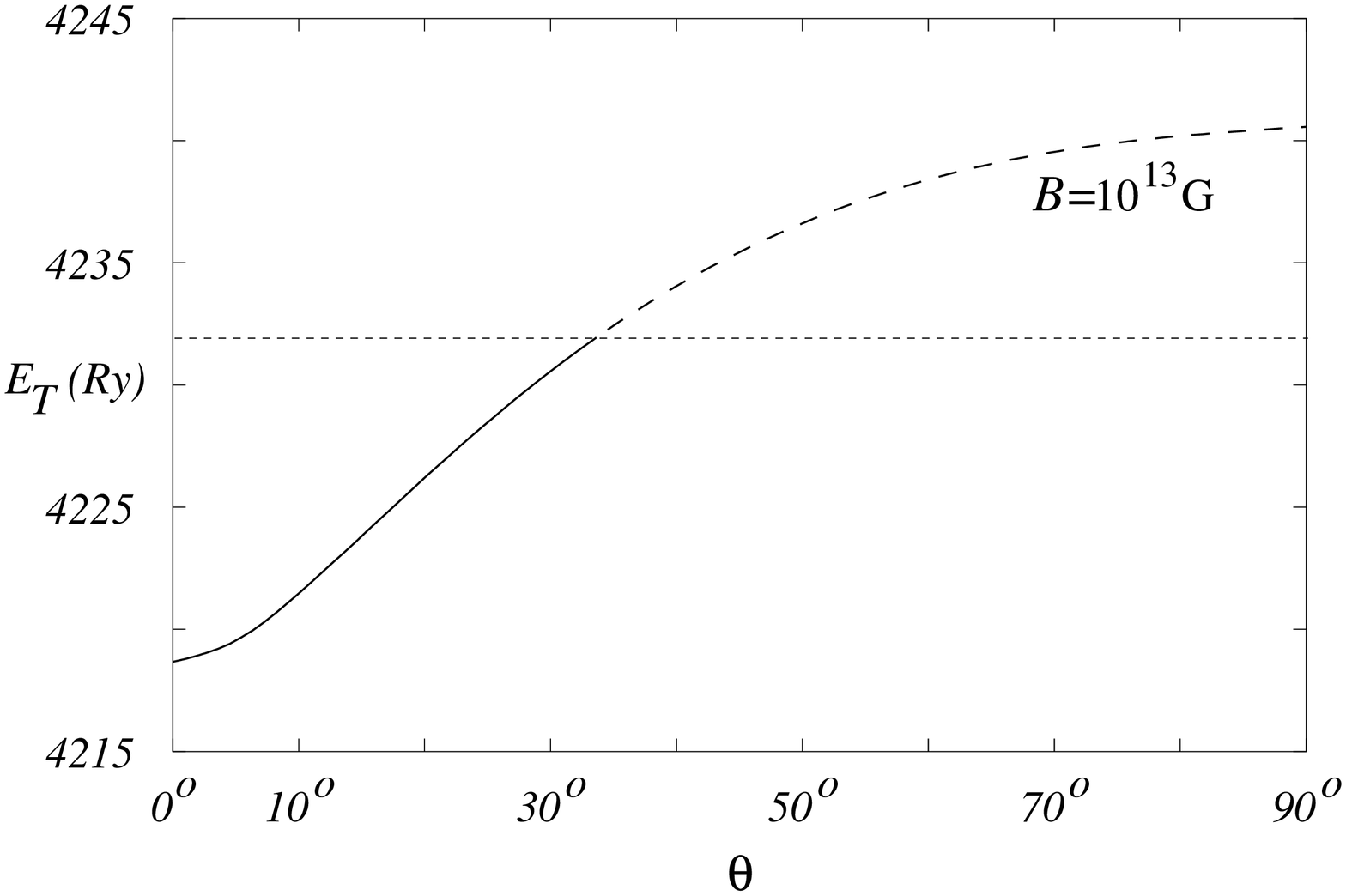}}     \\
    {\includegraphics[width=2.4in,angle=-90]{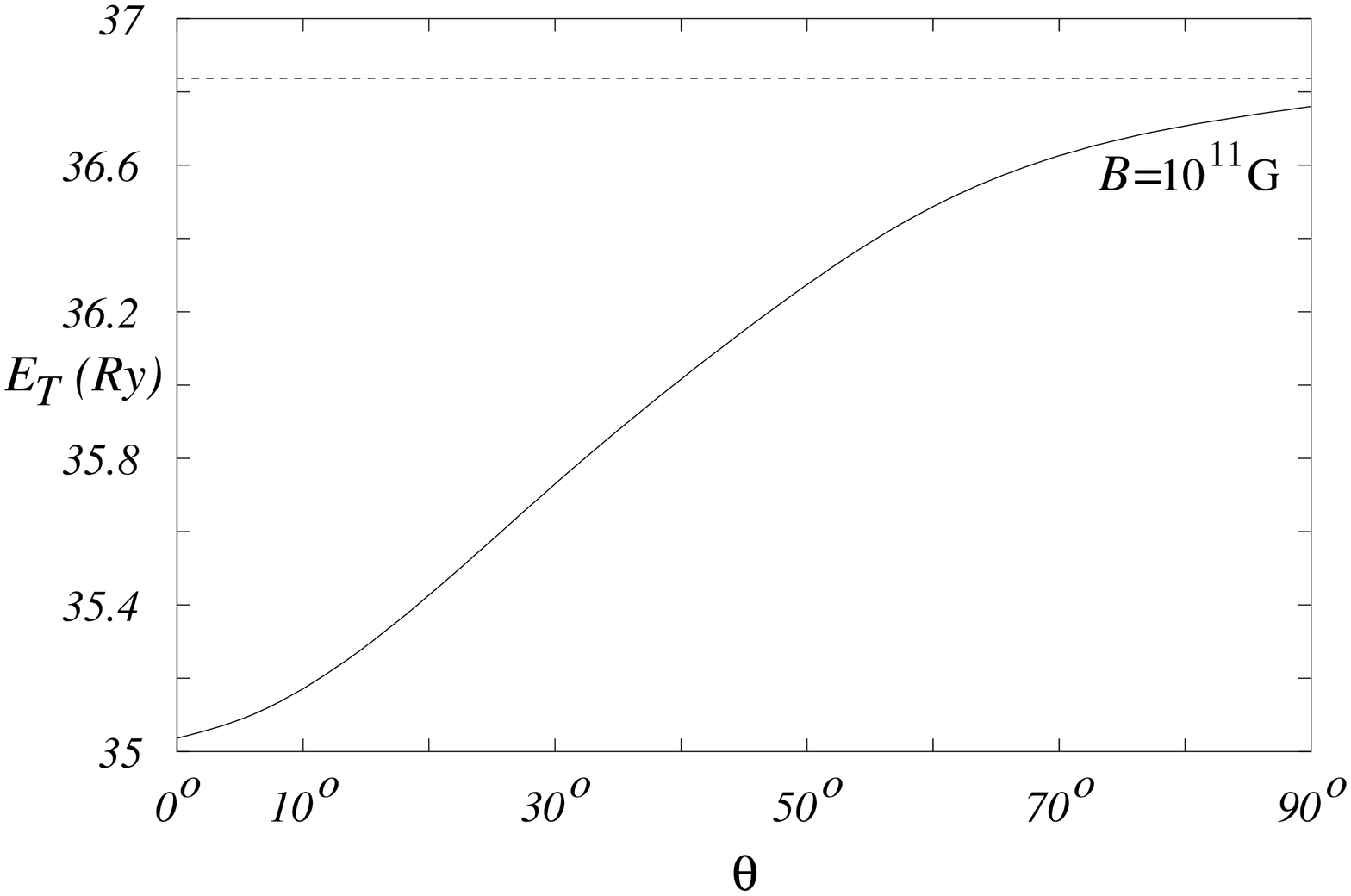}}&
    {\includegraphics[width=2.4in,angle=-90]{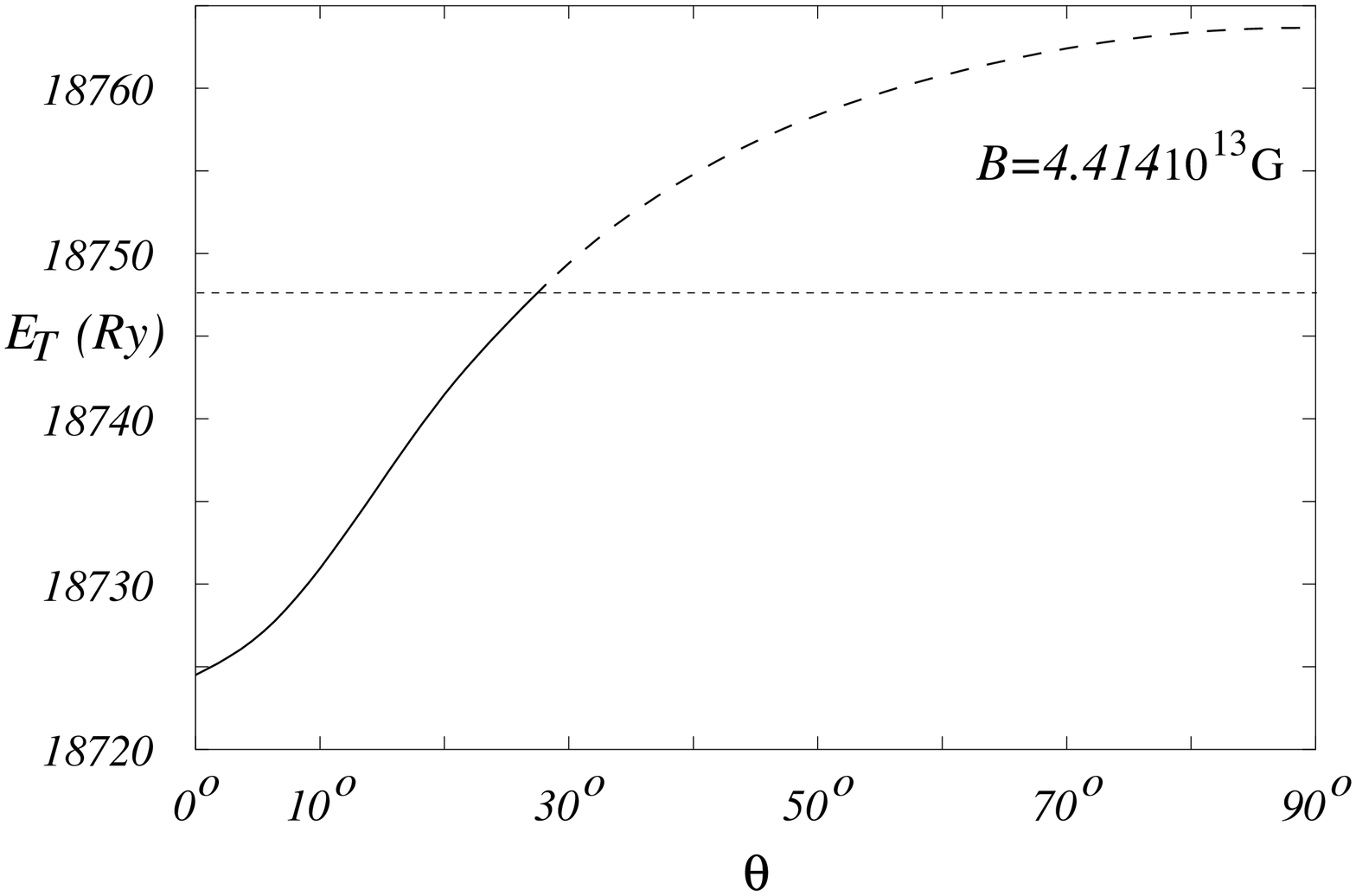}}
    \end{array}
     \]
    \caption{\label{fig:11} $H_2^+$ total energy ($E_T$) for the ground state
      $1_g$ as function of the inclination angle $\tha$ for different
      magnetic fields. The dotted lines correspond to the H-atom total
      energy taken from \cite{Salpeter:1992}.
      Dashed lines describe a total energy corresponding to a spurious minimum
      (see discussion in the text).}
  \end{center}
\end{figure*}

In a similar way the binding energy $E_b=B - E_T$, as well as the
dissociation energy (affinity to a hydrogen atom) $E_d=E_{H} -
E_T$ as a function of $\tha$ always decreases when changing from
the parallel to the perpendicular configuration (see Fig. 11).
Such behavior holds for all values of the magnetic field strength
studied. Thus, we can draw the conclusion that the molecular ion
becomes less and less stable monotonically as a function of
inclination angle. This confirms the statement made in \cite{Kher,
Wille:1988, Larsen, Schmelcher}, that the {\it highest molecular
stability of the $1_g$ state of $H_2^+$ occurs for the parallel
configuration}. Thus, {\it the $H_2^+$ molecular ion is the most
stable in parallel configuration}.

We extend the validity of this statement to magnetic field
strengths $10^{13} < B \lesssim 4.414 \times 10^{13}\,G$. It is
worth emphasizing that the rate of increase of binding energy
$E_b$ with magnetic field growth depends on the inclination -- it
slows down with  increased inclination. This effect implies that
the $H_2^+$-ion in the parallel configuration becomes more and
more stable against rotations -- the energy of the lowest
rotational state increases rapidly with magnetic field (see Table
V below and the discussion there).

Regarding the interproton equilibrium distance $R_{eq}$, one would
naively expect that it would always decrease with inclination (see
Fig. 12). Indeed, for all the magnetic fields studied we observe
that $R_{eq}$ at $\tha=0^{\circ}$ is larger than for any $\tha
\neq 0^{\circ}$ (see below, Tabs. I, II, III). This can be
explained as a natural consequence of the much more drastic
shrinking of the electronic cloud in the direction transverse to
the magnetic field than in the longitudinal one. Actually, for
magnetic fields $B \lesssim 10^{12}\,G$ the equilibrium distance
$R_{eq}$ decreases monotonically with inclination growth until it
reaches $\tha_{cr}$, as seen in Fig. 12. As mentioned above, if
the parameters of the vector potential (\ref{Vec}) are kept fixed,
$\xi=1/2$ and $Y=Z=0\, (d=0)$, a spurious minimum appears and
generates anomalous (spurious) $R_{eq}$  behavior for $\tha >
\tha_{cr}$(see \cite{Turbiner:2001}).

\begin{figure*}
  \begin{center}
    \[
    \begin{array}{cc}
     \includegraphics[width=2.4in,angle=-90]{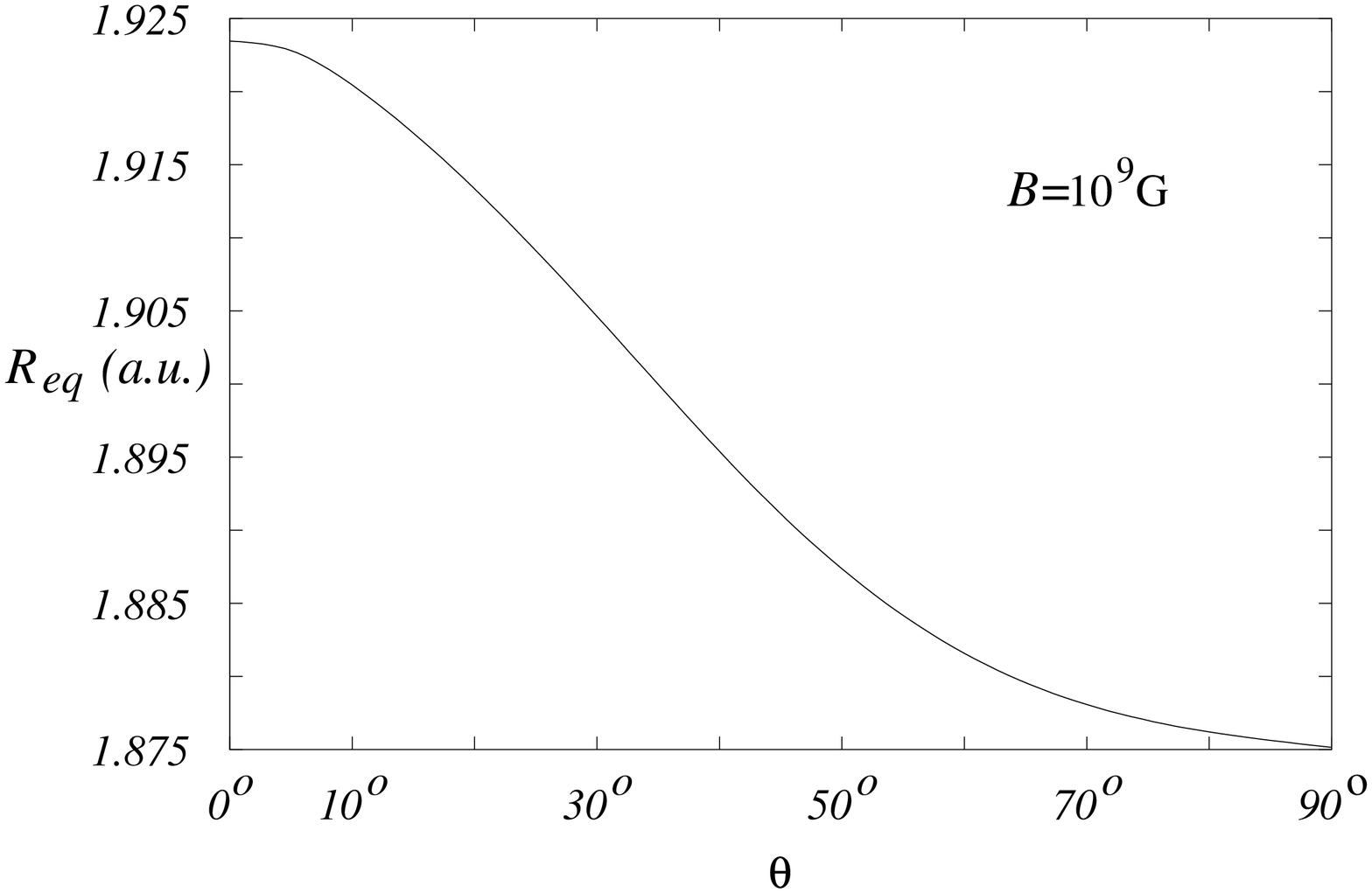} &
     \includegraphics[width=2.4in,angle=-90]{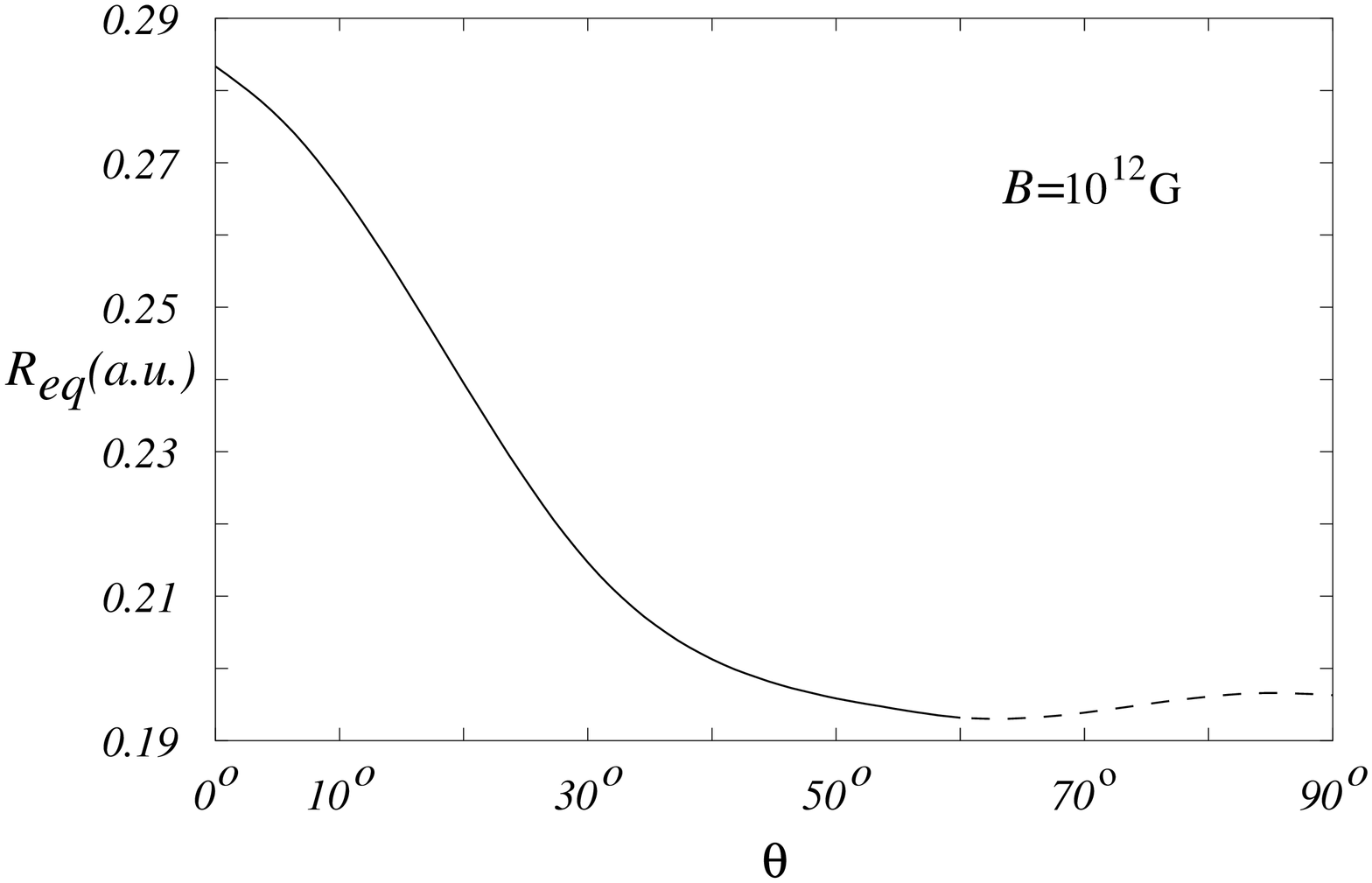} \\
     \includegraphics[width=2.4in,angle=-90]{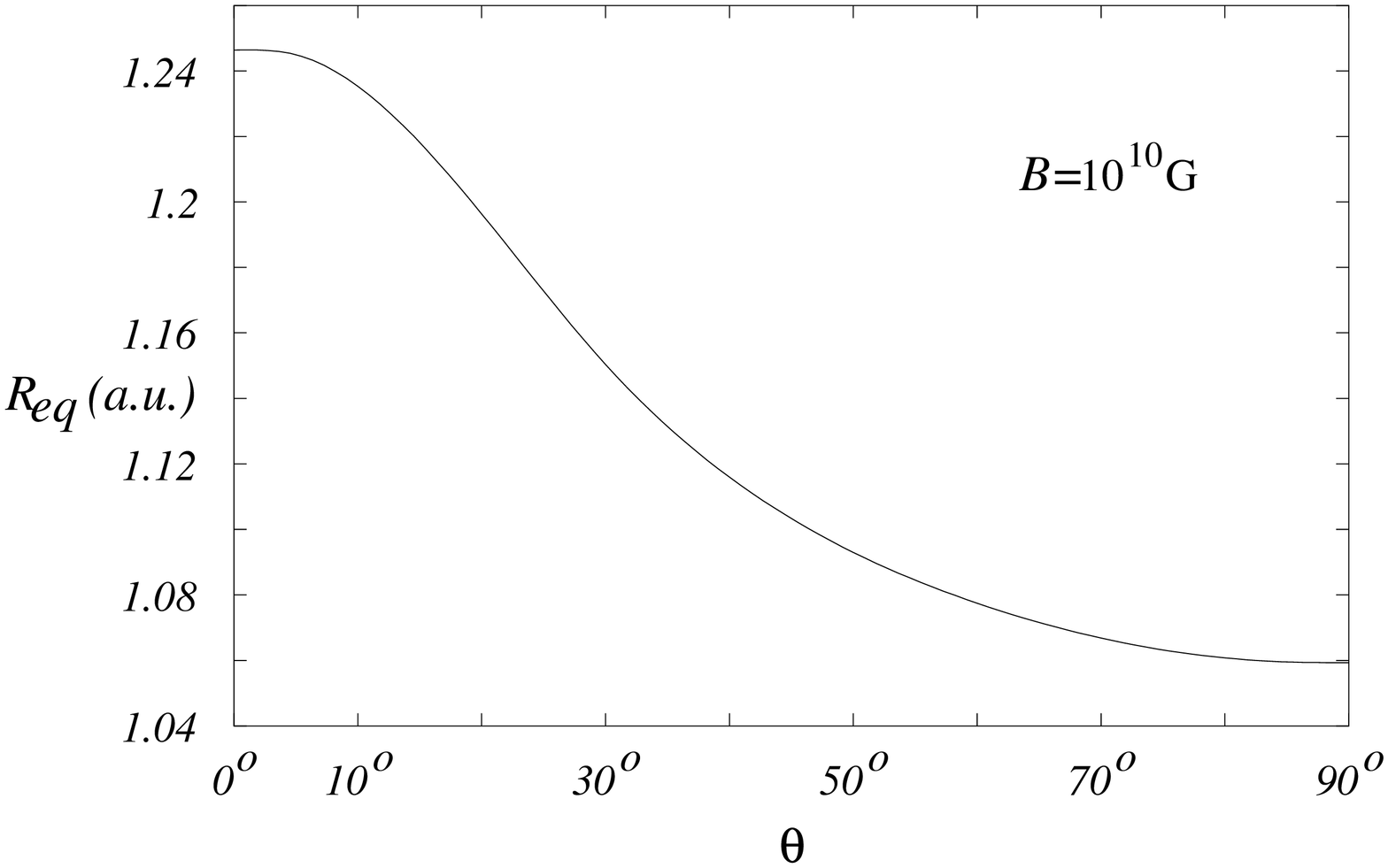}&
     \includegraphics[width=2.4in,angle=-90]{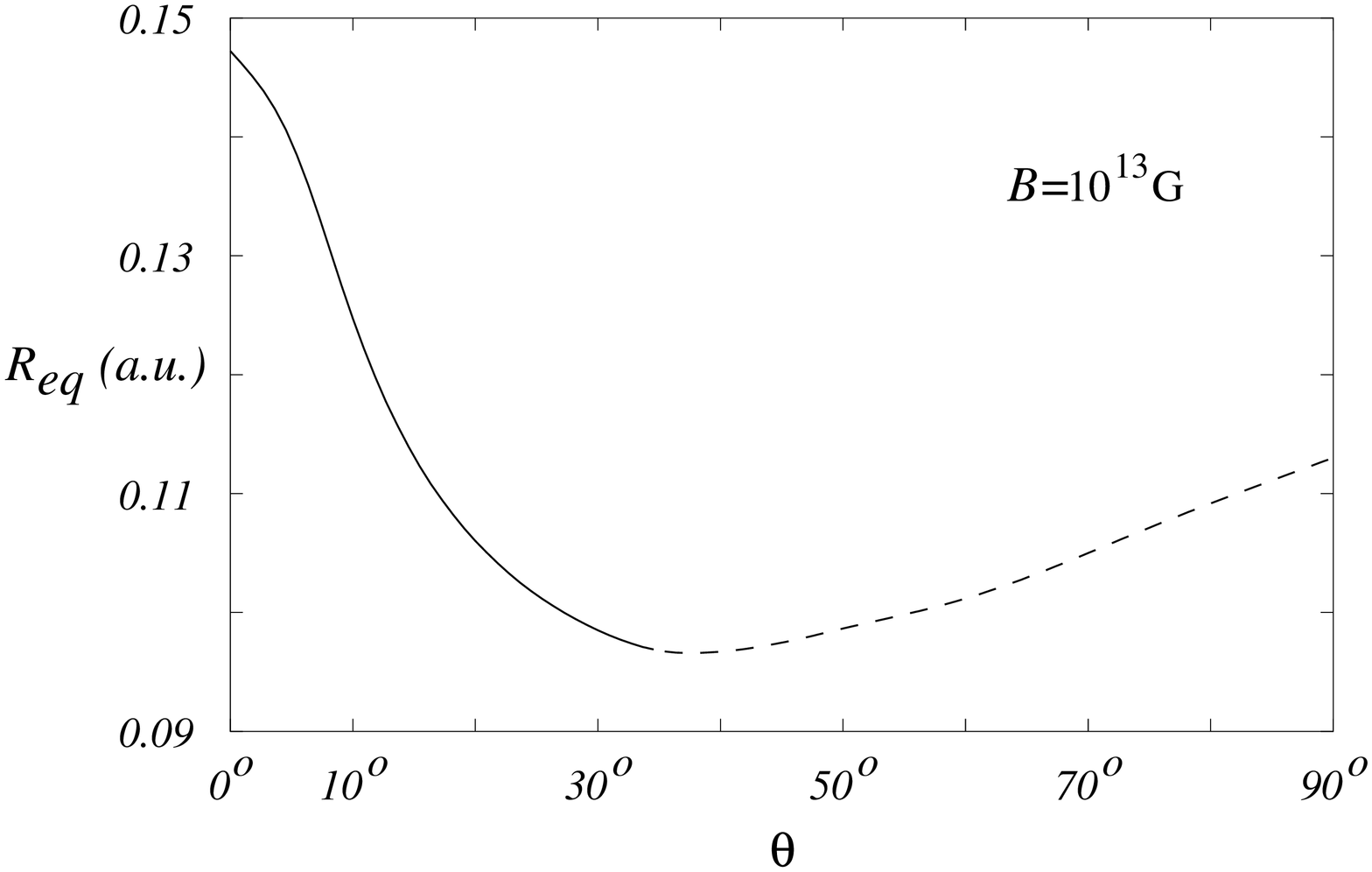}\\
     \includegraphics[width=2.4in,angle=-90]{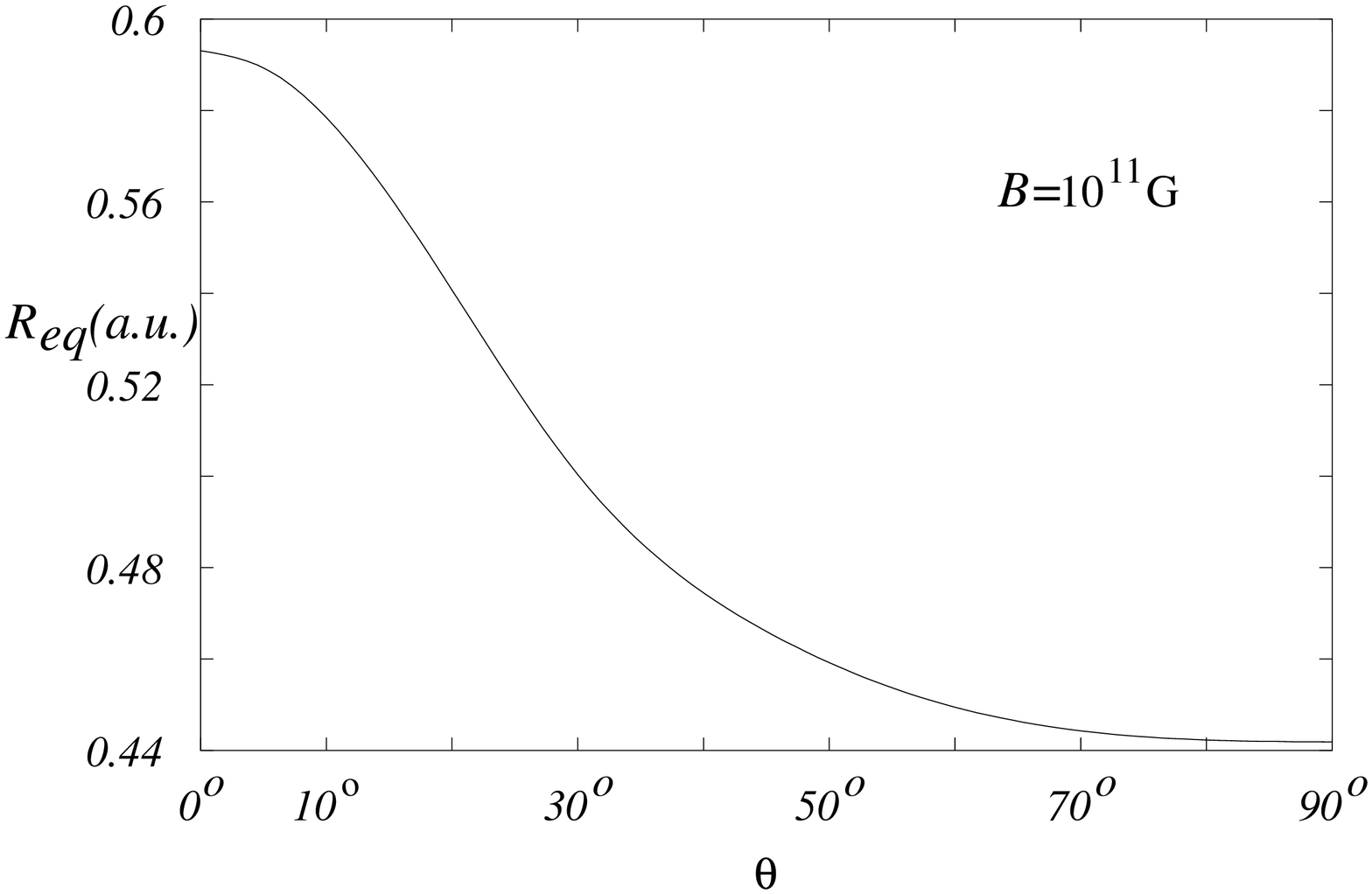}&
     \includegraphics[width=2.4in,angle=-90]{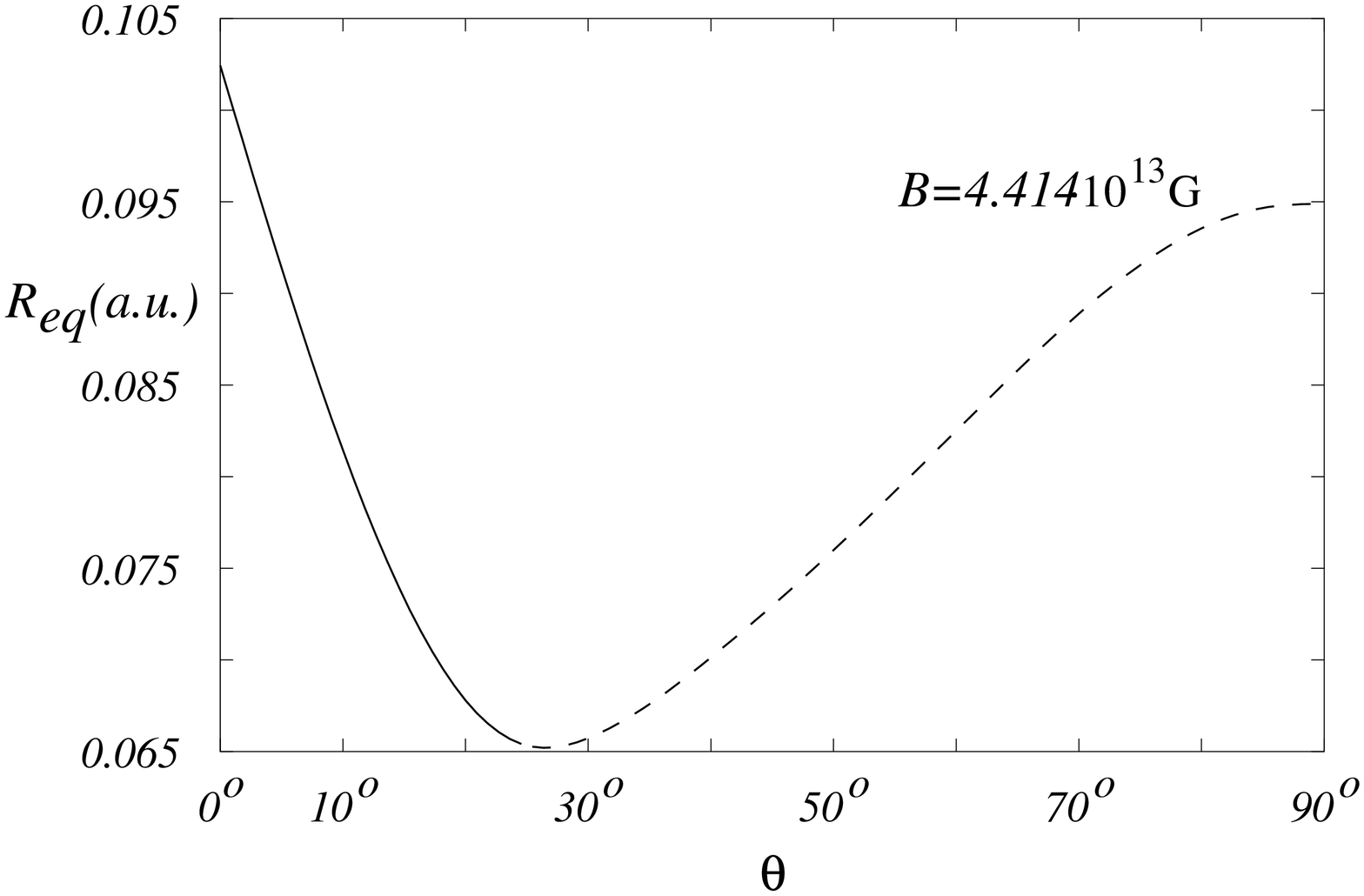}
    \end{array}
     \]
    \caption{\label{fig:12} $H_2^+$ equilibrium  distance as
    function of the inclination angle $\tha$ for the $1_g$ state.
    Dashed lines describe the position of a spurious minimum
    (see discussion in the text and Fig. 11).}
  \end{center}
\end{figure*}

In Tabs. I, II and III the numerical results for the total energy
$E_T$, binding energy $E_b$ and equilibrium distance $R_{eq}$ are
displayed for $\tha$=0$^{\circ}$, $45^{\circ}$ and $90^{\circ}$,
respectively. As seen in Table I, our results for $\tha=0^{\circ}$
lead to the largest binding energies for $B > 10^{11}\,G$ in
comparison with previous calculations. For $B \lesssim
10^{11}\,G$, our binding energies for the parallel configuration
appear to be very close (of the order of $\lesssim 10^{-4} -
10^{-5}$ in relative deviation) to the variational results by
Wille \cite{Wille:1988}, which are the most accurate so far in
this region of magnetic field strengths. Those results are based
on the use of a trial function in the form of a linear
superposition of $\sim 500$ Hylleraas type functions. It is quite
striking that our simple trial function (8) with ten variational
parameters gives comparable (for $B \lesssim 10^{11}\,G$) or even
better (for $B > 10^{11}\,G$) accuracy. It is important to reveal
the reason why the trial function \cite{Wille:1988} fails to be
increasingly inaccurate with magnetic field growth for
$B>10^{11}\,G$. An explanation of this inaccuracy is related to
the fact that in the $(x,y)$- directions the exact wave function
decays asymptotically as a Gaussian function, unlike the Hylleraas
functions which decay as the exponential of a linear function. The
potential corresponding to the function \cite{Wille:1988}
reproduces correctly the original potential near Coulomb
singularities but fails to reproduce $\sim \rho^2$-growth at large
distances. This implies a zero radius of convergence of the
perturbation theory for which the variational energy represents
the first two terms (see discussion in \cite{Tur}).

\begingroup
\squeezetable
\begin{table*}
\caption{\label{Table:1}
  Total, $E_T$, binding, $E_b$, energies and equilibrium distance $R_{eq}$
  for the state $1_g$ in the parallel configuration,  $\tha=0^{\circ}$.
  ${}^{\dagger}$~This value is taken from \cite{Lopez:1997}}
\begin{ruledtabular}
\begin{tabular}{lcccl}
    $B$ &  $E_T$ (Ry) &  $E_{b}$ (Ry) & $R_{eq}$ (a.u.) &  \\
    \hline
  $ B=0 $
          & -1.20525  & ---    & 1.9971 &  Present${}^{\dagger}$\\
          & -1.20527  & ---    & 1.997  &  Wille \cite{Wille:1988}\\
  $ 10^{9}\,G$
          & -1.15070  & 1.57623    & 1.924 &  Present\\
          & -1.15072  & 1.57625    & 1.924  &  Wille \cite{Wille:1988}\\
  $ 1\,a.u.$
          & -0.94991  & 1.94991    & 1.752  &  Present\\
          & ---       & 1.9498     & 1.752  &  Larsen \cite{Larsen}\\
          & -0.94642  & 1.94642    & 1.76   &  Kappes et al \cite{Schmelcher}\\
  $ 10^{10}\,G$
          & 1.09044   &  3.16488   & 1.246  &  Present\\
          & 1.09031   &  3.16502   & 1.246  &  Wille \cite{Wille:1988}\\
  $ 10\,a.u.$
          & 5.65024   &  4.34976   & 0.957  & Present\\
          & ---       &  4.35      & 0.950  & Wille \cite{Wille:1988}\\
          & ---       &  4.35      & 0.958  & Larsen \cite{Larsen}\\
          & ---       &  4.3346    & 0.950  & Vincke et al \cite{Vincke}\\
  $ 10^{11}\,G$
          & 35.0434   & 7.50975    & 0.593  &  Present\\
          & 35.0428   & 7.5104     & 0.593  &  Wille \cite{Wille:1988}\\
          & ---       & 7.34559    & 0.61   &  Lai et al \cite{Salpeter:1992} \\
  $ 100\,a.u.$
          & 89.7090   & 10.2904    & 0.448  & Present\\
          & ---       & 10.2892    & 0.446  & Wille \cite{Wille:1988}\\
          & ---       & 10.1577    & 0.455  & Wunner et al \cite{Wunner:82}\\
          & ---       & 10.270     & 0.448  & Larsen \cite{Larsen}\\
          & ---       & 10.2778    & 0.446  & Vincke et al \cite{Vincke}\\
  $ 10^{12}\,G$
          & 408.3894  & 17.1425    & 0.283  & Present\\
          & ---       & 17.0588    & 0.28   & Lai et al \cite{Salpeter:1992}\\
          & 408.566   & 16.966     & 0.278  & Wille \cite{Wille:1988}\\
  $ 1000\,a.u$
          & 977.2219  & 22.7781    & 0.220  & Present\\
          & ---       & 21.6688    & 0.219  & Wille \cite{Wille:1988}\\
          & ---       & 22.7069    & 0.221  & Wunner et al \cite{Wunner:82}\\
          & ---       & 22.67      & 0.222  & Larsen \cite{Larsen}\\
          & ---       & 22.7694    & 0.219  & Vincke et al \cite{Vincke}\\
  $ 10^{13}\,G$
          & 4219.565  & 35.7539    & 0.147  & Present\\
          & 4231.82   & 23.52      & 0.125  & Wille \cite{Wille:1988}\\
          & ---       & 35.74      & 0.15   & Lai et al \cite{Salpeter:1992}\\
  $ 4.414\times{10}^{13}\,G$
          & 18728.48  & 54.4992    & 0.101  & Present\\
\end{tabular}
\end{ruledtabular}
\end{table*}
\endgroup

The results for $\tha=45^{\circ}$ are shown in Table II, where a
gradual shortening of the equilibrium distance is accompanied by
an increase of total and binding energies with magnetic field. It
is worth noting that the parameter $\xi$ evolves from about $0.5$
to $0.93$ with magnetic field growth, thus changing from the
symmetric gauge for weak fields to an almost asymmetric one for
strong ones. This phenomenon takes place for all orientations $0
<\tha < \tha_{cr}$, becoming more and more pronounced with
increasing inclination angle (see below). We are unaware of any
other calculations for $\tha=45^{\circ}$ to compare ours with.

\begin{table}
 \caption{\label{Table:2} Total $E_T$, binding $E_b$ energies  and
  equilibrium distance  $R_{eq}$ for the $1_g$ state  at
  $\tha=45^{\circ}$. Optimal value of the gauge parameter $\xi$
  is given and $d=0$ is assumed (see text).}
\begin{ruledtabular}
\begin{tabular}{lcccc}
\( B \)& \( E_T \) (Ry) &\(    E_{b} \) (Ry) &\( R_{eq} \) (a.u.)&
\( \xi \)\\[5pt] \hline
 \( 10^{9} \) G
              & -1.14248  & 1.56801  & 1.891 & 0.5806  \\[5pt]
 \( 1 \) a.u.
              & -0.918494  & 1.918494  & 1.667 & 0.5855  \\[5pt]
\( 10^{10}\) G
              & 1.26195  & 2.99337 & 1.103 & 0.5958   \\[5pt]
 \( 10 \) a.u.
              & 6.02330  & 3.97670 & 0.812 & 0.6044   \\[5pt]
 \( 10^{11}\) G
              & 36.15633  & 6.39686 & 0.466 & 0.6252   \\[5pt]
 \( 100 \) a.u.
              & 91.70480  & 8.29520 & 0.337 & 0.6424   \\[5pt]
 \( 10^{12}\) G
              & 413.2987  & 12.2332 & 0.198 & 0.6890   \\[5pt]
 \( 1000\) a.u.
              & 985.1956  & 14.8044 & 0.147 & 0.7151   \\[5pt]
\end{tabular}
\end{ruledtabular}
\end{table}

For the perpendicular configuration $\tha = 90^{\circ}$, the
results are presented in Table III. Similar to what appeared for
the parallel configuration (see above) our results are again
slightly less accurate than those of Wille \cite{Wille:1988} for
$B \lesssim 10^{10}\,G$, but becoming the most accurate results
for stronger fields. In particular, it indicates that the domain
of applicability of a trial function in the form of a
superposition of Hylleraas type functions, becomes smaller as the
inclination grows. The results reported by Larsen \cite{Larsen}
and by Kappes-Schmelcher \cite{Schmelcher} are slightly worse than
ours, although the difference is very small.  The evolution of the
gauge parameters follow a similar trend, as was observed at $\tha
= 45^{\circ}$. In particular, $\xi$ varies from $\xi=0.64$ to
$\xi=0.98$ with magnetic field growth from $B=10^9\,G$ to $B \sim
2 \times 10^{11}\,G$ \footnote{$\xi=0.5$ at $B=0$}. We should
emphasize that the results of Larsen \cite{Larsen} and Wille
\cite{Wille:1988} for $B > 10^{11}\,G$ do not seem relevant
because of loss of accuracy, since the $H_2^+$ ion does not exist
in this region.

\begin{table*}
\caption{\label{Table:3}
 Total, $E_T$, and binding, $E_b$, energies and the
  equilibrium distance  $R_{eq}$ for the  $1_g$ state in the
  perpendicular configuration, $\tha=90^{\circ}$. The
  optimal value of the gauge parameter $\xi$ is presented and $d$
  is kept fixed, $d=0$ (see text).}
\begin{ruledtabular}
 \begin{tabular}{lccccl}
\(B\)&  \( E_T \) (Ry)& \(E_{b}\) (Ry)& \(R_{eq}\) (a.u.)&
\(\xi\)&
\\
\hline \( 10^{9}\) G
          & -1.137342 & 1.56287  & 1.875  & 0.6380 & Present\\
          &           & 1.56384  & 1.879  &        & Wille
           \cite{Wille:1988}\\
  \( 1 \) a.u.
          & -0.89911  & 1.89911  & 1.635  & 0.6455 & Present\\
          & ---       & 1.8988   & 1.634  &        & Larsen \cite{Larsen}\\
          & -0.89774  & 1.8977   & 1.65   &        & Kappes et al
           \cite{Schmelcher}\\
  \( 10^{10}\) G
          &  1.36207  & 2.89324  & 1.059  & 0.6621 & Present\\
          &  ---      & 2.8992   & 1.067  &        & Wille
           \cite{Wille:1988} \\
  \( 10\) a.u.
          &  6.23170  & 3.76830  & 0.772  & 0.6752 & Present\\
          & \emph{---}& 3.7620   & 0.772  &        & Larsen \cite{Larsen}\\
  \(10^{11} \) G
          & 36.7687   & 5.78445  & 0.442  & 0.7063 & Present\\
          & ---       & 5.6818   & 0.428  &        & Wille
           \cite{Wille:1988}\\
  \( 100\) a.u.
          & 92.7346   & 7.26543  & 0.320  & 0.7329 & Present\\
          & ---       & 7.229    & 0.320  &        & Larsen
           \cite{Larsen}\\
  \( 10^{12}\) G
          & --  & -- & -- & -- & Present\\
          & --- & 4.558  & 0.148 &        & Wille
           \cite{Wille:1988}\\
  \( 1000\) a.u.
          & --  & -- & -- & -- & Present\\
          & --- & 11.58 & 0.1578 &     & Larsen
           \cite{Larsen}\\
\end{tabular}
\end{ruledtabular}
\end{table*}

\begin{table}
 \caption{ \label{Table:4} The $1_g$ state: Expectation values of the
  transverse $<\rho>$ and longitudinal $<|z|>$ sizes of the electron
  distribution for the $H_2^+$-ion in a.u. at different orientations
  and magnetic field strengths.  At $\tha = 0^{\circ}$ the expectation
  value $<\rho>$ almost  coincides to the cyclotron radius of the electron.}
\begin{ruledtabular}
\begin{tabular}{lcccccc}
\( B \) & \multicolumn{3}{c}{\( <\rho > \) }& \multicolumn{3}{c}{
\( <|z|> \) }\\ & \( 0^{o} \)& \( 45^{o} \)& \( 90^{o} \)& \(
0^{o} \)& \( 45^{o} \)& \( 90^{o} \)\\ \hline \( 10^{9} \) G&
0.909 & 1.002& 1.084 & 1.666& 1.440 & 1.180 \\
\( 1 \) a.u.&
 0.801 & 0.866 & 0.929 & 1.534 & 1.313 & 1.090\\
\( 10^{10} \) G&
 0.511 & 0.538 & 0.569 & 1.144 & 0.972 & 0.848\\
\( 10 \) a.u.&
 0.359 & 0.375 & 0.396 & 0.918& 0.787 & 0.708\\
\( 10^{11} \) G&
 0.185 & 0.193 & 0.205 & 0.624 & 0.542 & 0.514\\
\( 10 \)0 a.u.&
 0.123 & 0.129 & 0.139 & 0.499 & 0.443 & 0.431 \\
\( 10^{12} \) G&
 0.060 & 0.065 & --    & 0.351 & 0.324 & -- \\
\( 10 \)00 a.u.&
 0.039 & 0.043 & --    & 0.289 & 0.275 & -- \\
\( 10^{13} \) G&
 0.019 & -- & -- & 0.215 & -- & -- \\
\( 4.414\times 10^{13} \) G&
 0.009& -- & -- & 0.164 & -- & -- \\
\end{tabular}
\end{ruledtabular}
\end{table}

In order to characterize the electronic distribution of $H_2^+$
for different orientations we have calculated the expectation
values of the transverse $<\rho>$ and longitudinal $<|z|>$ sizes
of the electronic cloud (see Table IV). Their ratio is always
limited,
\[
\frac{<\rho>}{<|z|>} \ < \ 1\ ,
\]
and quickly decreases with magnetic field growth, especially for
small inclination angles. This reflects the fact that the
electronic cloud has a more and more pronounced needle-like form
oriented along the magnetic line, as was predicted in the
classical papers
[\onlinecite{Kadomtsev:1971}-\onlinecite{Ruderman:1971}]. The
behavior of $<\rho>$ itself does not display any unusual
properties, smoothly decreasing with magnetic field, quickly
approaching the cyclotron radius for small inclinations and large
magnetic fields. In turn, the $<|z|>$ monotonically decreases with
inclination growth.

As already mentioned, the results of our analysis of the parallel
configuration of $H_2^+$ turned out to be optimal for all magnetic
fields studied, being characterized by the smallest total energy.
Therefore, it makes sense to study the lowest vibrational and also
the lowest rotational state (see Table V). In order to do this we
separate the nuclear motion along the molecular axis near
equilibrium in the parallel configuration (vibrational motion) and
deviation in $\tha$ of the molecular axis from $\tha=0^{\circ}$
(rotational motion).  The vicinity of the minimum of the potential
surface $E(\tha, R)$ at $\tha = 0^{\circ}, R=R_{eq}$ is
approximated by a quadratic potential, and hence we arrive at a
two-dimensional harmonic oscillator problem in the $(R,
\tha)$-plane. Corresponding curvatures near the minimum define the
vibrational and rotational energies (for precise definitions and
discussion see, for example, \cite{Larsen}). We did not carry out
a detailed numerical analysis, making only rough estimates of the
order of $20 \%$. For example, at B$=10^{12}\,G$ we obtain
$E_{vib}= 0.276\,Ry$ in comparison with $E_{vib}=0.259\,Ry$ given
in \cite{Lopez-Tur:2000}, where a detailed variational analysis of
the potential electronic curves was performed.  Our estimates for
the energy, $E_{vib}$, of the lowest vibrational state are in
reasonable agreement with previous studies.  In particular, we
confirm a general trend of the considerable increase of
vibrational frequency with the growth of $B$ indicated for the
first time by Larsen \cite{Larsen}. The dependence of the energy
on the magnetic field is much more pronounced for the lowest
rotational state -- it grows much faster than the vibrational one
with magnetic field increase. This implies that the $H_2^+$-ion in
the parallel configuration becomes more stable for larger magnetic
fields (see the discussion above). From a quantitative point of
view the results obtained by different authors are not in good
agreement. It is worth mentioning that our results agree for large
magnetic fields $\gtrsim 10\,a.u.$ with results by Le Guillou et
al. \cite{Legui}, obtained in the framework of the so called
`improved static approximation', but deviate drastically at
$B=1\,a.u.\,$, being quite close to the results of Larsen
\cite{Larsen} and Wille \cite{Wille:1987}. As for the energy of
the lowest rotational state, our results are in good agreement
with those obtained by other authors (see Table V).

\begingroup
\squeezetable
\begin{table*}
  \caption{\label{Table:5}
    Energies of the lowest vibrational $(E_{vib})$ and rotational
    $(E_{rot})$ electronic states  associated with the $1_g$ state
    at $\tha=0^{\circ}$.  The indexes in Le Guillou et al
    \cite{Legui} are assigned to the `improved adiabatic approximation'
    (a), and to the `improved static approximation' (b). }
\begin{ruledtabular}
\begin{tabular}{lcccl}
\( B \)& \( E_T \) (Ry)& \( E_{vib} \) (Ry)& \( E_{rot} \) (Ry)&
\\ \hline
 \( 10^{9} \) G
              & -1.15070 & 0.013 & 0.0053 & Present \\
              &  ---     & 0.011 & 0.0038 & Wille
               \cite{Wille:1987} \\
\( 1 \) a.u.
              & -0.94991 & 0.015 & 0.0110 & Present \\
              &    ---   & ---   & 0.0086 & Wille
               \cite{Wille:1987} \\
              &    ---   & 0.014 & 0.0091 & Larsen
               \cite{Larsen} \\
              &    ---   & 0.013 & ---    & Le Guillou et al (a)
               \cite{Legui}\\
              &    ---   & 0.014 & 0.0238 & Le Guillou et al (b)
               \cite{Legui}\\
\( 10^{10} \) G
              & 1.09044  & 0.028 & 0.0408 & Present \\
              &   ---    & 0.026 & 0.0308 & Wille \cite{Wille:1987}\\
\( 10 \) a.u.
              & 5.65024  & 0.045 & 0.0790 & Present \\
              &    ---   & 0.040 & 0.133  & Larsen\cite{Larsen} \\
              &    ---   & 0.039 &  ---   & Le Guillou et al (a)
              \cite{Legui}\\
              &    ---   & 0.040 & 0.0844 & Le Guillou et al (b)
              \cite{Legui}\\
\( 10^{11} \) G
              & 35.0434  & 0.087 & 0.2151 & Present \\
\( 100 \) a.u.
              & 89.7096  & 0.133 & 0.4128 & Present \\
              &  ---     & 0.141 & 0.365  & Larsen\cite{Larsen}\\
              &  ---     & 0.13  &  ---   & Wunner et al
               \cite{Wunner:82}\\
              &    ---   & 0.128 &  ---   & Le Guillou et al (a)
               \cite{Legui}\\
              &    ---   & 0.132 & 0.410  & Le Guillou et al (b)
               \cite{Legui}\\
\( 10^{12} \) G
              & 408.389  & 0.276 & 1.0926 & Present \\
              &   ---    & 0.198 & 1.0375 & Khersonskij
               \cite{Kher3} \\
\( 1000 \) a.u.
              & 977.222  & 0.402 & 1.9273 & Present \\
              &   ---    & 0.38  & 1.77   &  Larsen\cite{Larsen} \\
              &   ---    & 0.39  &  ---   & Wunner et al
               \cite{Wunner:82}\\
              &    ---   & 0.366 &  ---   & Le Guillou et al (a)
               \cite{Legui}\\
              &    ---   & 0.388 & 1.916  & Le Guillou et al (b)
               \cite{Legui}\\
\( 10^{13} \) G
              & 4219.565 & 0.717 & 4.875  & Present \\
              &   ---    & 0.592 & 6.890  & Khersonskij
               \cite{Kher3}  \\
\( 4.414\times 10^{13} \)G
              & 18728.48 & 1.249 & 12.065 & Present\\
\end{tabular}
\end{ruledtabular}
\end{table*}
\endgroup

In Fig. 13 we show the electronic distributions $\int dy
|\psi(x,y,z)|^2$ for magnetic fields $10^{9}, 10^{10}, 10^{11},
10^{12}\,G$ and different orientations for $H_2^+$ in the
equilibrium configuration, $R=R_{eq}$. It was already found
explicitly \cite{Lopez:1997} that at $\tha=0^o$ with magnetic
field increase there is a change from `ionic' (two-peak electronic
distribution) to `covalent' coupling (one-peak distribution). We
find that a similar phenomenon holds for all inclinations. If for
$B=10^9\,G$, all electronic distributions are characterized by two
peaks for all inclinations, then for $B=10^{12}\,G$ all
distributions have a  single sharp peak. The `sharpness' of the
peak grows with magnetic field. Fig. 13 also demonstrates how the
change of the type of coupling appears for different inclinations
-- for larger inclinations a transition $(two-peaks)
\leftrightarrow (one-peak)$ appears for smaller magnetic fields.
It seems natural that for the perpendicular configuration
$\tha=90^{\circ}$, where the equilibrium distance is the smallest,
this change appears for even smaller magnetic field.

\begin{figure*}
  \begin{center}
    \[
     \]\vskip -10pt
    \caption{\label{fig:13} Electronic distributions
    $\int dy |\psi(x,y,z)|^2$  (normalized to one) for the $1_g$ state
    of $H_2^+$ (equilibrium configuration) for different magnetic fields
    and inclinations.}
  \end{center}
\end{figure*}

In Figs. 14-18 we present the evolution of the electronic
distributions as a function of interproton distance $R$, for
inclinations $0^o, 45^o$ at $B=1 a.u.$ and $10^{12}\,G$ together
with the $R$-dependence for the inclination $90^o$ at $B=1
a.u.\,$. The values of magnetic fields are chosen to illustrate in
the most explicit way the situation. In all figures a similar
picture is seen. Namely, at not very large magnetic fields $B
\lesssim 10^{11}\,G$ and for all inclinations the electronic
distribution at small $R < R_{cr}$ is permutationally symmetric
and evolves with increase of $R$ from a one-peak to a two-peak
picture with more and more clearly pronounced, separated peaks.
Then for $R=R_{cr}$ this symmetry becomes broken and the electron
randomly chooses one of protons and prefers to stay in its
vicinity. For $R \gg R_{cr}$ the electronic distribution becomes
totally asymmetric, the electron looses its memory of the second
proton. This signals that the chemical reaction $H_2^+ \rar H + p$
has already happened. For larger magnetic fields $B \gtrsim
10^{11}\,G$ for $R < R_{cr}$ the electronic distribution is always
single-peaked, a transition from a one-peak to a two-peak picture
occurs for $R > R_{cr}$, where the electronic distribution is
already asymmetric.

\begin{figure}[htbp]
   \begin{center}
       \[
     \begin{array}{cc}
\end{array}
\]
  \caption{\label{fig:14} Evolution of the electronic distributions
    $\int dy |\psi(x,y,z)|^2$  (normalized to one) and their contours for
    the $1_g$ state of the $(ppe)$ system with interproton distance for
    $B = 1\, a.u.,\, \tha=0^o$.}
   \end{center}
\end{figure}

\begin{figure}[htbp]
   \begin{center}
       \[
     \begin{array}{cc}
\end{array}
\]
 \caption{\label{fig:15} Evolution of the electronic distributions
    $\int dy |\psi(x,y,z)|^2$  (normalized to one) and their contours for
    the $1_g$ state of the $(ppe)$ system with interproton distance for
    $B = 1\, a.u.,\, \tha=45^o$.}
   \end{center}
\end{figure}

\begin{figure}[htbp]
   \begin{center}
       \[
     \begin{array}{cc}
\end{array}
\]
 \caption{\label{fig:16} Evolution of the electronic distributions
    $\int dy |\psi(x,y,z)|^2$  (normalized to one) and their contours for
    the $1_g$ state of the $(ppe)$ system with interproton distance for
    $B = 1\, a.u., \tha=90^o$.}
   \end{center}
\end{figure}

\begin{figure}[htbp]
   \begin{center}
       \[
     \begin{array}{cc}
\end{array}
\]
 \caption{\label{fig:17} Evolution of the electronic distributions
    $\int dy |\psi(x,y,z)|^2$  (normalized to one) and their contours for
    the $1_g$ state of the $(ppe)$ system with interproton distance for
    $B = 10^{12}\,G, \tha=0^o$.}
   \end{center}
\end{figure}

\begin{figure}[htbp]
   \begin{center}
       \[
     \begin{array}{cc}
\end{array}
\]
 \caption{\label{fig:18} Evolution of the electronic distributions
    $\int dy |\psi(x,y,z)|^2$  (normalized to one) and their contours for
    the $1_g$ state of the $(ppe)$ system with interproton distance for
    $B = 10^{12}\,G, \tha=45^o$.}
   \end{center}
\end{figure}

To complete the study of the $1_g$ state we show in Fig. 19 the
behavior of the variational parameters of the trial function
(\ref{trial}) as a function of the magnetic field strength for the
optimal (parallel) configuration, $\tha=0^o$. In general, the
behavior of the parameters is rather smooth and {\it very}
slowly-changing, even though the magnetic field changes by several
orders of magnitude. This is in drastic contrast with the results
of Kappes-Schmelcher \cite{Kappes:1994} (see Fig. 1 in this
paper). In our opinion such behavior of the parameters of our
trial function (9) reflects the level of adequacy (or, in other
words, indicates the quality) of the trial function. In practice,
the parameters can be approximated by the spline method and then
can be used to study magnetic field strengths other than those
presented here.

\begin{figure}
 \begin{center}
  \[
  \begin{array}{c}
  \begin{picture}(215,160)(-10,0)
  \put(5,20){$10^9$} \put(40,20){$10^{10}$} \put(83,20){$10^{11}$}
  \put(125,20){$10^{12}$} \put(165,20){$10^{13}$} \put(85,8){$B
  (G)$}
  \put(-10,32){$-4$} \put(-10,62){$-3$} \put(-10,92){$-2$}
  \put(-10,122){$-1$} \put(-10,152){$\,\,\, 0$}
  \put(150,120){$A_2$} \put(110,90){$A_1$}
  \put(-10,165){\includegraphics*[width=2.3in,angle=-90]{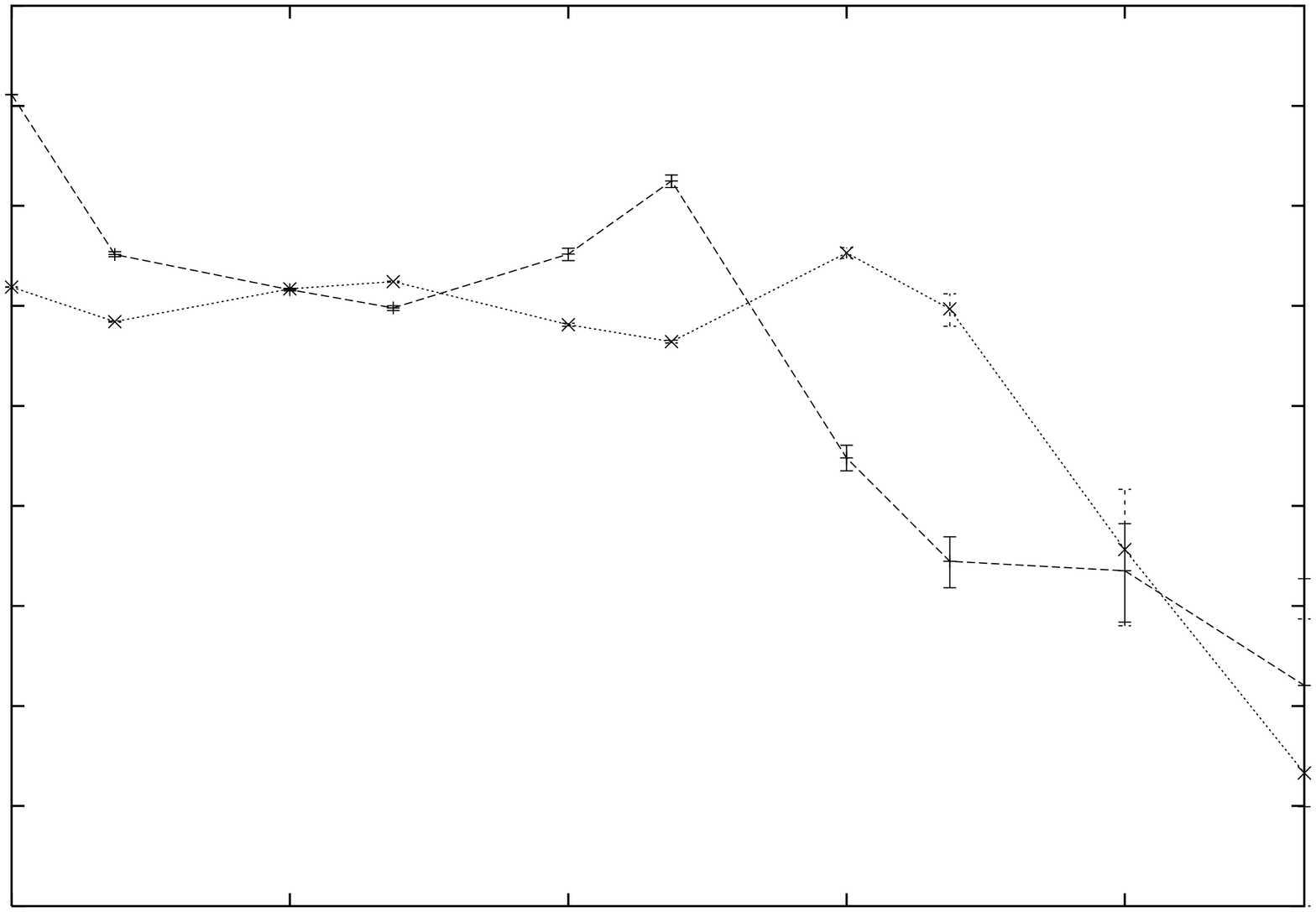}}
  \end{picture}
\\[-20pt]
  \begin{picture}(215,263)
  \put(10,20){$10^9$} \put(45,20){$10^{10}$} \put(88,20){$10^{11}$}
  \put(130,20){$10^{12}$} \put(170,20){$10^{13}$} \put(90,8){$B
  (G)$}
  \put(5,35){$0$} \put(5,85){$4$} \put(5,130){$8$} \put(4,180){$12$}
  \put(4,225){$16$} \put(-22,155){$[a.u.]^{-1}$}
  \put(170,180){$\alpha_3$} \put(160,130){$\alpha_2$}
  \put(150,80){$\alpha_1$} \put(140,42){$\alpha_4$}
  \put(-22,0){\includegraphics*[width=3.35in,height=3.65in]{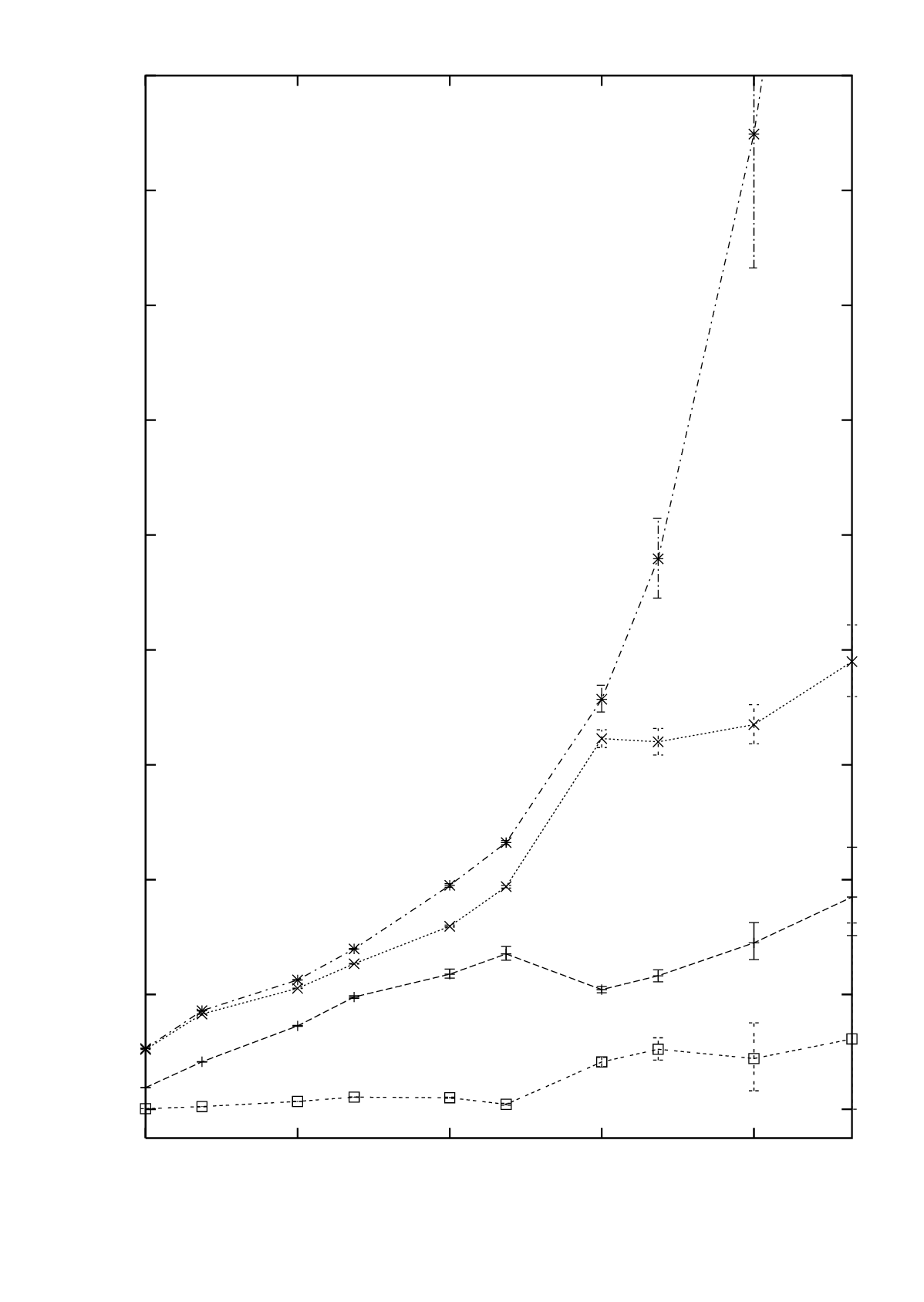}}
  \end{picture}
\\[-20pt]
  \begin{picture}(215,180)(-10,0)
  \put(5,20){$10^9$} \put(40,20){$10^{10}$} \put(83,20){$10^{11}$}
  \put(125,20){$10^{12}$} \put(165,20){$10^{13}$} \put(85,8){$B
  (G)$}
  \put(-12,42){$0.2$} \put(-12,65){$0.4$} \put(-12,87){$0.6$}
  \put(-12,110){$0.8$} \put(-12,132){$1.0$} \put(-12,152){$1.2$}
  \put(-35,97){$[a.u.]^{-1}$}
  \put(80,105){$\beta_{2}$} \put(70,136){$\beta_{3}$}
  \put(65,60){$\beta_{1}$}
  \put(-10,167){\includegraphics*[width=2.3in,angle=-90]{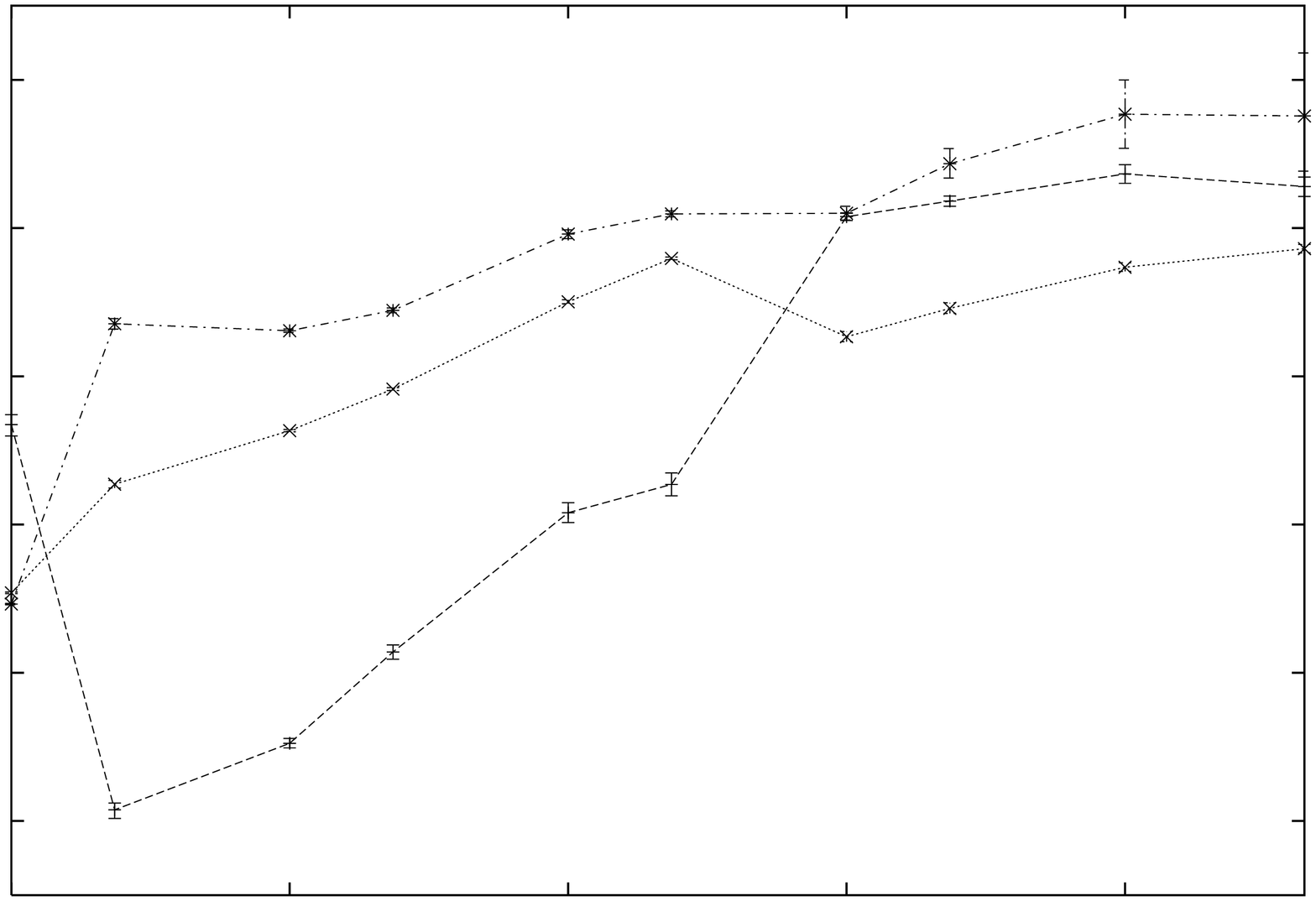}}
  \end{picture}
  \end{array}
\] \vskip -20pt
 \caption{Variational parameters of the trial function (9) as a
  function of the magnetic field strength $B$ for the $1_g$ state
  in the parallel configuration, $\tha=0^{\circ}$. In this case
  $\beta_1 = \frac{\beta_{1x}}{2}=\frac{\beta_{1y}}{2}\,,\,
  \beta_2 = \frac{\beta_{2x}}{2}=\frac{\beta_{2y}}{2}\,,\,
  \beta_3 = \frac{\beta_{3x}}{2}=\frac{\beta_{3y}}{2}$
  (see eqs.(5,6,8), cf. \cite{Lopez:1997} ).
  The parameter $A_3$  is fixed, being equal to $1$ and $\xi=1/2, d=0$
  (see text).
  The error bars correspond to relative deviation in the variational
  energy in the region
  $\Delta E_{T}\equiv\frac{E_T}{E_{var}}\lesssim~10^{-5}$.}
\end{center}
\end{figure}

\newpage

\section{Conclusion}

We have carried out an accurate, non-relativistic calculation in
the Born-Oppenheimer approximation for the lowest state of the
$H_2^+$ molecular ion for different orientations of the magnetic
field with respect to the molecular axis. We studied constant
uniform magnetic fields ranging from zero up to $B = 4.414 \times
10^{13}\,G$, where non-relativistic consideration holds.

For all magnetic fields studied there exist a region of
inclinations for which a well-pronounced minimum in the total
energy surface for the $1_g$ state of the system $(ppe)$ is found.
This shows the existence of the $H_2^+$ molecular ion for magnetic
fields $B = 0 - 4.414 \times 10^{13}\,G$. The smallest total
energy is always found to correspond to the parallel
configuration, $\tha=0^{\circ}$, where protons are situated along
the magnetic line. The total energy increases, while the binding
energy decreases monotonically as the inclination angle grows. The
rate of total energy increase as well as binding energy decrease
is seen to be always maximal for the parallel configuration. The
equilibrium distance exhibits quite natural behavior as a function
of the orientation angle $\tha$ -- for fixed magnetic field the
shorter equilibrium distance always corresponds to the larger
$\tha$.

Confirming the qualitative observations made by Khersonskij
\cite{Kher} for the $1_g$ state in the contrast to statements in
\cite{Larsen, Wille:1988}, we demonstrate accurately that the
$H_2^+$-ion does not exist at a certain range of orientations for
magnetic fields $B \gtrsim 2 \times 10^{11}\,G$. As the magnetic
field increases the region of inclinations where $H_2^+$ does not
exist is seen to broaden, reaching rather large domain
$25^{\circ}\lesssim \tha \leqslant 90^{\circ}$ for $B=4.414 \times
10^{13}\,G$.

We find that the electronic distributions for $H_2^+$ in the
equilibrium position are qualitatively different for weak and
large magnetic fields. In the domain $B < 10^{10}\,G$ the
electronic distribution for any inclination has a two-peak form,
peaking near the position of each proton. On the contrary for
$B>10^{11}\,G$ the electronic distribution always has a single
peak form with the peak near the midpoint between the protons for
any inclination. This implies a physically different structure for
the ground state - for weak fields the ground state can be
modelled as a `superposition' of hydrogen atom and proton, while
for strong fields such modelling is not appropriate.

Unlike standard potential curves for molecular systems in the
field-free case, we observe for $\tha > 0^o$ that each curve has a
maximum and approaches to the asymptotics at $R \rar \infty$ from
above. The electronic distribution evolves with $R$ from a
one-peak form at small $R$ to a two-peak one at large $R$. There
exists a certain critical $R_{cr}$ at which one of peaks starts to
diminish, manifesting a breaking of permutation symmetry between
the protons and simultaneously the beginning of the chemical
reaction $H_2^+ \rar H + p$.

Combining all the above-mentioned observations we conclude that
for magnetic fields of the order of magnitude $B \sim 10^{11}\,G$
some qualitative changes in the behavior of the $H_2^+$ ion take
place. The behavior of the variational parameters also favors this
conclusion. This hints at the appearance of a new scale in the
problem. It might be interpreted as a signal of a transition to
the domain of developed quantum chaos (see, for example,
\cite{chaos}).

\begin{acknowledgments}
The authors wish to thank B.I.~Ivlev and M.I.~Eides for useful
conversations and interest in the subject. We thank M. Ryan for a
careful reading of the manuscript.

This work was supported in part by DGAPA Grant \# IN120199
(M\'exico) and CONACyT Grant 36650-E (M\'exico). AT thanks the
University Program FENOMEC for financial support.
\end{acknowledgments}


\end{document}